\documentclass[11pt,a4paper]{article}
\pdfoutput=1
\usepackage{jheppub}
\usepackage[utf8]{inputenc}
\usepackage{amsfonts}
\usepackage{amsmath}
\usepackage{color}
\usepackage[normalem]{ulem}
\usepackage{verbatim}
\usepackage{subcaption}
\usepackage{braket}

\title{Colour Reconnection from Soft Gluon Evolution}
\author[a]{Stefan Gieseke}
\author[a]{Patrick Kirchgae\ss er}
\author[b]{Simon Pl\"atzer}
\author[c,d]{Andrzej Siodmok}
\affiliation[a]{Institute for Theoretical Physics, Karlsruhe Institute of
  Technology}
\affiliation[b]{Particle Physics, Faculty of Physics, University of Vienna}
\affiliation[c]{The Henryk Niewodniczanski Institute of Nuclear Physics in
  Cracow}
\affiliation[d]{Czech Technical University in Prague, Brehova 7, 115 19 Prague, Czech Republic}

\emailAdd{stefan.gieseke@kit.edu}
\emailAdd{patrick.kirchgaesser@kit.edu}
\emailAdd{simon.plaetzer@univie.ac.at}
\emailAdd{andrzej.siodmok@ifj.edu.pl}

\preprint{\begin{flushright}KA-TP-23-2018\\UWTHPH-2018-23\\HERWIG-2018-02\\MCnet-18-22\end{flushright}}

\abstract{We consider soft gluon evolution at the amplitude level to expose
  the structure of colour reconnection from a perturbative point of
  view. Considering the cluster hadronization model and an universal Ansatz
  for the soft anomalous dimension we find strong support for geometric models
  considered earlier. We also show how reconnection into baryonic systems
  arises, and how larger cluster systems evolve. Our results provide the
  dynamic basis for a new class of colour reconnection models for cluster
  hadronization.}

\begin{document}

\maketitle

\section{Introduction}

Multi-purpose Monte Carlo event generators
(MCEG)~\cite{Bahr:2008pv,Sjostrand:2006za,Sjostrand:2014zea,Gleisberg:2008ta}
play a central role for experimental analyses and phenomenological
investigations by simulating particle collisions at a realistic level,
including the full complexity and several orders of magnitude difference in
relevant energy scales. Starting from a so-called hard scattering at a large
momentum transfer, and possibly multiple partonic interactions in hardonic
collisions, subsequent radiation in a parton shower is building up the first
level of jet substructure and as such provides the first and foremost input to
predicting the physical behaviour of observables. These multiple emissions do
account for leading logarithmic contributions at all orders of the strong
coupling, which can be systematically addressed within analytic approaches to
the re-summation of the QCD perturbative series either by the direct analysis
of scattering amplitudes and cross sections, or by means of effective field
theory methods.

As the typical scales or inter-parton separations reach small scales of the
order of $1-2\ {\rm GeV}$, perturbative evolution needs to stop and
phenomenological models are used to describe the transition of the partonic
ensemble into the observed hadrons. These models take into account
non-perturbative corrections, which are included in the analytical approach by
means of non-perturbative shape functions, and typically interpret low-mass
partonic systems as excited hadronic systems which then break up into stable
or unstable hadrons. This applies both to the string
hadronization~\cite{Andersson:1983ia, Fischer:2016zzs} as well as cluster
hadronization paradigms \cite{Webber:1983if} employed in LHC-age MC event
generators. The physical picture behind these models is that at the end of the
perturbative evolution colour charge is already confined into small phase space
regions such that interpretation of excited hadronic systems does make sense,
however in the complex and dense environment of hadron-hadron collisions this
picture can be spoiled, and is typically invalidated by the presence of
multiple partonic scatters.

One therefore expects that further dynamics of exchanging colour charges
between the systems are present, which will reduce the relative separations of
colour singlet systems. These physics is encoded in colour reconnection
models, which are crucial for the description of minimum bias and underlying
event data
\cite{Sjostrand:1993hi,Lonnblad:1995yk,Gieseke:2012ft,Christiansen:2015yca,
  Bierlich:2015rha,Reichelt:2017hts} at hadron colliders. The lack of most
models to describe baryon production at hadron colliders has also led to
improved models of colour reconnection
\cite{Christiansen:2015yqa,Gieseke:2017clv}.

While the spectrum of colour singlet systems is typically already predicted by
the large-$N$ parton shower evolution, the colour reconnection dynamics is a
genuine sub-leading-$N$ effect which we expect to be composed of by a
perturbative as well as a non-perturbative component. Though the effects have
yet only been addressed within the context of non-perturbative models, work is
ongoing in analysing multi-parton emission dynamics beyond the leading-$N$
approximation \cite{Platzer:2013fha,Martinez:2018ffw} stemming from a detailed
analysis of factorisation properties of cross sections, and the resummation of
logarithmically enhanced terms. While no consistent connection of such
approaches to established hadronization models has been made yet, we take --
in this work -- the perturbative structure as a starting point to calculate
colour reconnecting effects in amplitude level evolution as an input to more
constrained, and hence more predictive colour reconnection models.

This paper is organised as follows: In Sec.\,\ref{sec:Preconfinement} we
review the details of colour pre-confinement and cluster
hadronization. Following this in Sec.\,\ref{sec:ColourEvolution} we introduce
the concept of perturbative colour evolution of a scattering amplitude,
detailing how different colour structures get mixed in the (infrared)
renormalised amplitude by means of a renormalization group equation in colour
space. Using this as a proxy to design how a perturbatively inspired colour
reconnection model would look like, we state a general algorithm in
Sec.\,\ref{sec:GeneralAlgorithm}, where we calculate the probabilities to
change from an initial colour flow to a final one by evaluating overlaps of
the evolved amplitude and a target colour structure, including baryonic
configurations. Sec.\,\ref{sec:TwoClusterSandbox} analyses in very detail the
(exactly solvable) evolution of a two-cluster system in various kinematic
regimes and provides an analytic back-up of the dynamics which are typically
considered in phenomenological approaches, before we present numerical results
for systems of small clusters using a full colour flow evolution in
Sec.\,\ref{sec:NumericalResults}. These results are then considered in
Sec.\,\ref{sec:FullModel} to isolate building blocks for a full-fledged new
model of colour reconnection, the implementation of which is subject to
ongoing work.

\section{Preconfinement and Cluster Hadronization}
\label{sec:Preconfinement}

The cluster hadronization model\,\cite{WEBBER1984492} is an essential
ingredient for Monte Carlo Event Generators such as
Herwig\,\cite{Bellm:2015jjp} and Sherpa\,\cite{Gleisberg:2008ta} to convert
the partons at scales of the parton shower infrared cutoff of order $1\ {\rm
  GeV}$ into observed hadrons at energy scales of order
$\Lambda_{\text{QCD}}$.  The cluster model is based on the property of colour
preconfinement\,\cite{AMATI197987} which essentially states that at any scale
the colour structure of the parton shower is such that colour singlet
combinations of partons can be formed with an asymptotically invariant mass
distribution and that this mass distribution is independent of the properties
of the hard scattering process or the parton shower itself.

The colour flow of an event is determined through the parton shower, which
generally uses the leading colour approximation in order to define the colour
flow of a splitting. Since in the large $N$ limit\,\cite{HOOFT1974461} the
singlet contribution from emitting gluons is colour suppressed, they can be
represented through a colour and an anti-colour line, unambiguously determining
the colour flow of a splitting in the parton shower evolution of a state.  At
the end of the parton shower evolution each coloured parton is colour
connected to an anti-coloured parton forming a colour singlet cluster.  The
properties of a cluster are uniquely defined by the invariant cluster mass
\begin{equation}
\label{eq:invMass}
 M^2 = \left(p_{q} +p_{\bar{q}} \right)^2 \ ,
\end{equation}
and the kinematics and the flavours of the consitutent quarks ($q$, $\bar{q}$)
forming the cluster. Since our analysis will primarily concentrate on the
cluster level the specific flavours of the quarks are not of interest and can
be neglected.  Also for our purposes it is sufficient to stay in the massless
parton limit in which Eq.\,\ref{eq:invMass} simplifies to $M^2=2p_{q}\cdot
p_{\bar{q}}$.  Mass effects of light quarks are briefly discussed in
Sec.\,\ref{sec:Parameters} but not investigated further.

The assignment of colour connections between quark and anti-quark pairs is not
without flaws. While at $e^+e^-$ collisions the colour connections emerging
from the parton shower do lead to an asymptotically invariant mass
distribution of clusters, the situation becomes ill defined when multiple
parton interactions, as they appear during hadronic collisions, are taken into
account. Since it is unclear how the colour connections between different
scattering centres emerges, non-perturbative models are necessary in order to
rearrange the colour flow to arrive at a sensible description of
data\,\cite{Sjostrand:1987su}. One paradigm which inspired at least the
development of one model is the so called notion of a colour pre-confined
state which states that after the evolution of the parton shower has
terminated, the colour connected partons are close in momentum space leading
to a distribution of invariant cluster masses which peaks at small values
dictated by the parton shower infrared cutoff. Current developments in this
direction, regarding space-time hadronization models are underway and seem to
look promising for future studies\,\cite{Ferreres-Sole:2018vgo}.

While the bulk of the developments in recent years has been focused more on
the non-perturbative modelling side of event generation there hasn't been much
progress on finding further motivation for colour reconnection or
rearrangements of colour flows from the perturbative point of view.  In this
paper we approach this vast topic with a perturbatively inspired evolution of
the colour flow due to soft-gluon exchanges and analyse the properties of the
resulting cluster configurations which are favoured by our perturbative
{\it Ansatz}.

\section{Perturbative Colour Evolution}
\label{sec:ColourEvolution}

QCD scattering amplitudes are vectors in both colour and spin space,
and can be decomposed in a basis of contributing colour structures,
\begin{equation}
|{\cal M}\rangle = \sum_{\sigma} {\cal M}_\sigma |\sigma\rangle \ ,
\end{equation}
where we have suppressed the helicity degrees of freedom as we are
mainly interested in the colour structure. These bases are typically
over-complete, and non-orthogonal. This poses a computational, but not
a conceptual constraint, and in this work we consider the colour flow
basis, which at first sight has the worst scaling behaviour in terms
of over-completeness, however also exhibits the closest link to the
actual flow of colour charge through a scattering amplitude, and
provides us with a convenient connection to the parton shower
evolution and (pre-)confinement properties. We stress that this basis
is not limited to considerations in the large-$N$ limit, and has
indeed been shown to be a convenient tool for organising full-colour
evolution at the amplitude level
\cite{Platzer:2013fha,Martinez:2018ffw}, besides its earlier use in
the efficient recursive computation of tree-level amplitudes, see {\it
  e.g.}  \cite{Maltoni:2002mq}. Specifically, colour structures in the
colour flow basis can be labelled by permutations which describe how
colour charge is flowing from one leg to another,
\begin{equation}
|\sigma\rangle =\left|
\begin{array}{ccc}
1 & \cdots & n \\
\sigma(1) &\cdots & \sigma(n)
\end{array}
\right\rangle  = \delta^{\alpha_1}_{\bar{\alpha}_{\sigma(1)}}\cdots
\delta^{\alpha_n}_{\bar{\alpha}_{\sigma(n)}} \ .
\end{equation}  
Virtual corrections are in general both, ultraviolet and infrared
divergent. These divergencies are regulated within dimensional
regularization, and absorbed into renormalizing bare quantities at a
given scale $\mu^2$. While ultraviolet divergences in this way relate
to the running of the strong coupling, the infrared singularities
drive the evolution of a scattering amplitude and the renormalization
program in this case can be used to sum large logarithmic
contributions of infrared origin to all orders in perturbation
theory. To be precise, we can relate the bare amplitude
$|\tilde{\mathcal{M}}\rangle$ to the renormalized amplitude
$|\mathcal{M}\rangle$ as
\begin{equation}
  \label{eq:renAmplitude}
  |\mathcal{M}(\{p \}, \mu^2)\rangle =
  {\mathbf Z}^{-1}(\{p \},\mu^2,\epsilon) |\tilde{\mathcal{M}}(\{ p\},\epsilon)\rangle \ ,
\end{equation}
where $\{p\}$ is the set of outgoing momenta, $\epsilon = (d-4)/2$ is
the dimensional regularization parameter in $d$ dimensions, and
$\mu^2$ is the scale at which the (infrared) renormalization has been
performed. The renormalization constant ${\mathbf Z}$ is an operator
in the space of colour structures and sums the infrared divergences to
all orders, resulting in a finite renormalized amplitude.

By taking a logarithmic derivative of the bare amplitude with respect
to $\mu^2$ we obtain an evolution equation\footnote{We consider the
  evolution at most to one-loop level such that no implicit $\mu^2$
  dependence arises through the running of $\alpha_s$.}
\begin{equation}
  \mu^2 \frac{{\rm d}}{{\rm d}\mu^2} |{\cal M}(\{p\},\mu^2)\rangle =
    {\mathbf \Gamma}(\{p\},\mu^2)|{\cal M}(\{p\},\mu^2)\rangle
\end{equation}
where the soft anomalous dimension matrix
\begin{equation}
  {\mathbf \Gamma}(\{p\},\mu^2) = -{\mathbf
    Z}^{-1}(\{p\},\mu^2,\epsilon)\mu^2 \frac{\partial}{\partial \mu^2}{\mathbf Z}(\{p\},\mu^2,\epsilon)
\end{equation}
encodes the residues of the $1/\epsilon$ divergencies contained in
${\mathbf Z}$. At one loop, the soft anomalous dimension reads
\begin{equation}
  {\mathbf \Gamma}(\{p\},\mu^2) =\sum_{i\ne j}
  (-{\mathbf T}_i\cdot {\mathbf T}_j)
  \Gamma_{\text{cusp}}\ \ln\left(\frac{-s_{ij}}{\mu^2}\right) + \sum_i \gamma_i
\end{equation}
where $s_{ij}=2p_i\cdot p_j$ for two outgoing or two incoming massless
partons $i$ and $j$ and $s_{ij}=-2p_i\cdot p_j$ for incoming/outgoing
or outgoing/incoming pairs, and $\Gamma_{\text{cusp}}=\alpha_s/4\pi$
in lowest order. The solutions to the evolution equation take the
form
\begin{equation}
  \label{eq:evolSolution}
  |{\cal M}(\{p\},\mu^2)\rangle = {\mathbf U}(\{p\},\mu^2,\{M_{ij}^2\})
  |{\cal H}(\{p\}, Q^2,\{M_{ij}^2\})\rangle,
\end{equation}
where ${\cal{H}}(\{p\}, Q^2,\{M_{ij}^2\})$ represents the hard
scattering amplitude before the evolution.  
\begin{figure}
  \begin{center}
    \includegraphics[scale=0.5]{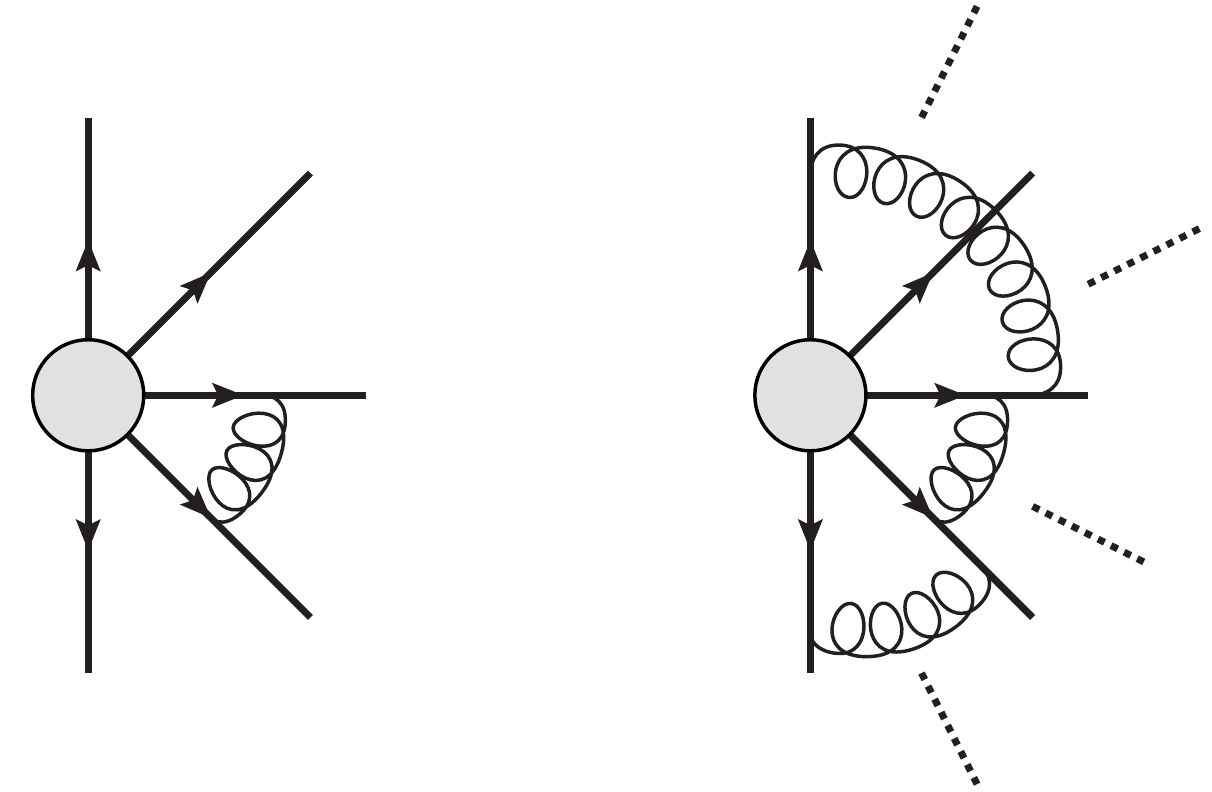}
  \end{center}
  \caption{\label{fig:gluonevolution}Diagrammatic representation of
    single soft gluon exchanges between two legs as described by the
    soft anomalous dimension matrix (left), and all possible iterated
    exchanges as encoded in the evolution operator (right). Thick lines
    indicate the momenta of the final state colour charges originating
    from the hard scattering indicated by the grey blob.}
\end{figure}
The evolution operator, neglecting the non-cusp terms $\gamma_i$ as
they are diagonal in colour space, and assuming that only final state,
massless partons are present, is
\begin{equation}
  \label{eq:leadingEvolution}
  {\mathbf U}\left(\{p\},\mu^2,\{M_{ij}^2\}\right) =
  \exp\left(-\sum_{i\ne j}\int_{\mu^2}^{M_{ij}^2}
  \frac{{\rm d}q^2}{q^2}(-{\mathbf T}_i\cdot {\mathbf T}_j)
  \Gamma_{\text{cusp}}\ \left(\ln\frac{2p_i\cdot p_j}{q^2} - i\pi\right) \right) \ .
\end{equation}
Here we have not chosen a fixed scale to provide the initial condition
for the evolution but rather have chosen an upper limit on the
integration per pair of partons, assuming that $\mu^2$ is always less
than the scales $M_{ij}^2$, which amounts to a specific choice of hard
amplitude which is already encoding logarithms of the universal hard
scale $Q^2$ and the specific mass parameters $M_{ij}^2$ which we
consider here. The reason for this splitting will soon become clear,
however the main point to make here is that the evolution operator
${\mathbf U}$ is a matrix exponential in colour space and is
describing iterated soft gluon exchanges between any two legs to all
orders in the strong coupling. We have illustrated this in
Fig.~\ref{fig:gluonevolution}.

In terms of the colour flow basis introduced earlier, the action of
the evolution operator can be summarised in iterating so-called colour
reconnectors \cite{Platzer:2013fha} which, once per action, will swap
two indices of the permutation labelling the specific colour flow, and
introduce longer sequences of transpositions when exponentiated. If
the colour flows in a basis tensor can be considered to indeed
represent physical colour singlet systems, than the evolution operator
can be expected to be the basic object describing the physics of
colour reconnection at the amplitude level. We shall use this
observation as a starting point for our model investigation.

\section{The General Algorithm and Baryonic Reconnections}
\label{sec:GeneralAlgorithm}

The specific configuration we obtain from a pre-confining parton
evolution as discussed in Sec.~\ref{sec:Preconfinement}, with a
universal cluster mass spectrum, can be seen as driven by a cross
section resulting from an amplitude which has been dominated by a
colour structure $|\tau\rangle$ corresponding to the assignment of
clusters identified in the final state,
\begin{equation}
  {\rm d}\sigma\sim |{\cal H}(\{p\}, Q^2,\{M_{ij}^2\})|^2\qquad
  |{\cal H}(\{p\}, Q^2,\{M_{ij}^2\})\rangle \approx {\cal H}_\tau(\{p\},
  Q^2,\{M_{ij}^2\}) |\tau\rangle  \ .
\end{equation}
In this case we assume that the logarithms of $Q^2/M_{ij}^2$ have been
summed by the parton shower evolution with $M_{ij}^2\sim 2p_i\cdot
p_j\sim Q_0^2$, effectively corresponding to a veto of radiation off
dipoles with masses around the shower infrared cutoff $Q_0^2$ and we
view the initial step of colour reconnection as an evolution in colour
space to scales of order $\mu^2$ below the initial cluster masses and
the parton shower infrared cutoff.  To this extend we identify
$M_{ij}^2 = 2p_i\cdot p_j$, and we use
\begin{equation}
\label{eq:EvolAnsatz}
{\mathbf U}\left(\{p\},\mu^2,\{M_{ij}^2\}\right) = 
   \exp\left( \sum_{i\ne j}
   {\mathbf T}_i\cdot {\mathbf T}_j
  \frac{\alpha_s}{2\pi} \left(\frac{1}{2}\ln^2\frac{M_{ij}^2}{\mu^2} - i\pi \ln\frac{M_{ij}^2}{\mu^2}\right) \right)
\end{equation}
as an Ansatz for the evolution.  The starting point for colour reconnection of
a cluster configuration represented through a colour structure $|\tau\rangle$
is then to consider the overlap between the evolved amplitude and a new colour
structure $|\sigma\rangle$ to constitute a {\it reconnection amplitude},
\begin{equation}
  {\cal A}_{\tau\to \sigma} = \langle \sigma |{\mathbf U}
  \left(\{p\},\mu^2,\{M_{ij}^2\}\right) | \tau \rangle \ .
\end{equation}
Here we have removed the partial amplitude for the colour flow we
start to evolve, as it will only an overall normalisation which is
irrelevant for the {\it reconnection probability}, which we now take
to be
\begin{equation}
  P_{\tau\to\sigma} = \frac{|{\cal A}_{\tau\to \sigma}|^2}{\sum_{\rho} |{\cal
      A}_{\tau\to \rho}|^2} \ ,
\end{equation}
where $\rho$ runs over all possible colour flows.
\subsection{Baryonic Colour Reconnections}
\label{sec:Baryons}

In \cite{Gieseke:2017clv} the concept of colour reconnection to
baryonic clusters has been investigated and proven to be central to
the description of baryon production at hadron colliders. In the
framework of our perturbatively inspired colour reconnection we can
also accommodate for such reconnections provided that there are at
least three clusters, or colour flows, to be considered. It is then
possible to associate a baryon/anti-baryon pair to a colour structure
which has been suitably anti-symmetrized in three fundamental and
three anti-fundamental indices
\begin{multline}
  |B_{ijk}\rangle = \frac{1}{N_B}\epsilon^{ijk}\epsilon_{\bar{i}\bar{j}\bar{k}} =\\
  \frac{1}{N_B}\left(
    \left|
\begin{array}{ccc}
i & j & k \\
\bar{i} &\bar{j} & \bar{k}
\end{array}
\right\rangle +
\left|
\begin{array}{ccc}
j & k & i \\
\bar{i} &\bar{j} & \bar{k}
\end{array}
\right\rangle +
\left|
\begin{array}{ccc}
k & i & j \\
\bar{i} &\bar{j} & \bar{k}
\end{array}
\right\rangle -
\left|
\begin{array}{ccc}
j & i & k \\
\bar{i} &\bar{j} & \bar{k}
\end{array}
\right\rangle -
\left|
\begin{array}{ccc}
i & k & j \\
\bar{i} &\bar{j} & \bar{k}
\end{array}
\right\rangle -
\left|
\begin{array}{ccc}
k & j & i \\
\bar{i} &\bar{j} & \bar{k}
\end{array}
\right\rangle
  \right) \ .
\end{multline}
The normalisation constant is taken to reproduce the normalisation of a single
mesonic configuration,
\begin{equation}
  \langle B_{ijk}|B_{ijk}\rangle = N^3 \ ,\qquad N_B^2 = 3!\left(1-\frac{3}{N} + \frac{2}{N^2}\right) =
  \frac{4}{3} \ .
\end{equation}
This allows us to define a {\it baryonic reconnection amplitude}
\begin{equation}
  {\cal A}_{\tau\to B_{ijk}\otimes \tilde{\sigma}_{ijk}} = \langle B_{ijk}|\otimes\langle
  \tilde{\sigma}_{ijk} | {\mathbf U}\left(\{p\},\mu^2,\{M_{ij}^2\}\right)|\tau\rangle \ ,
\end{equation}
where $\tilde{\sigma}_{ijk}$ denotes the permutation with the colour and
anti-colour indices corresponding to the baryonic system removed,
\begin{equation}
  |\tilde{\sigma}_{ijk}\rangle =
  \left|
\begin{array}{cccc}
1 & \cdots & n & \backslash\ i,j,k \\
\sigma(1) &\cdots & \sigma(n) & \backslash\ \bar{i},\bar{j},\bar{k}
\end{array}
\right\rangle \ .
\end{equation}
The generalised reconnection probability is then
\begin{equation}
  P_{\tau\to \sigma} = \frac{|{\cal A}_{\tau\to\sigma}|^2}{{\cal
      N}_{\tau}} \ ,\qquad
  P_{\tau\to B_{ijk}\otimes \tilde{\sigma}_{ijk}} = \frac{|{\cal A}_{\tau\to B_{ijk}\otimes \tilde{\sigma}_{ijk}}|^2}{{\cal
      N}_{\tau}}  
\end{equation}
with
\begin{equation}
  {\cal N}_{\tau} = \sum_{\rho} |{\cal
      A}_{\tau\to\rho}|^2 + \sum_\rho\sum_{i < j < k} | {\cal A}_{\tau\to
    B_{ijk}\otimes \tilde{\rho}_{ijk}}|^2 \ .
\end{equation}
We also consider the possibility of evolving an already existing Baryon, for
which we introduce 'unbaryonizing' reconnection amplitudes
\begin{equation}
  {\cal A}_{B_{ijk}\otimes \tilde{\sigma}_{ijk}\to \tau} = \langle\tau|
  {\mathbf U}\left(\{p\},\mu^2,\{M_{ij}^2\}\right)
  |B_{ijk}\rangle \otimes |\tilde{\sigma}_{ijk}\rangle \ .
\end{equation}
These allow us to quantify how relevant such an evolution step would
be for a high-mass baryonic system, which would not have entered the
reconnection dynamics any more in the case of the models considered
before.

\section{A Two-Cluster Sandbox}
\label{sec:TwoClusterSandbox}

The goal of this Section is to gain an analytical insight into colour
reconnection from soft gluon evolution. In order to do so we study the
simplest possible situation of the evolution of a two cluster system.  In this
case there are just two possible colour flows and according to
Eq.~\ref{eq:EvolAnsatz} the evolution of the single state can be expressed in
the following way:
\begin{equation}
\label{eq:twoclev}
   | \sigma\rangle = {\mathbf U} \left|
\begin{array}{cc}
i & j  \\
\bar{i} &\bar{j}
\end{array}
\right\rangle = e^{{\mathbf\Omega}} \left|
\begin{array}{cc}
i & j  \\
\bar{i} &\bar{j}
\end{array}
\right> = 
\sigma_{ij}\left|
\begin{array}{cc}
i & j  \\
\bar{i} &\bar{j}
\end{array}
\right> + \sigma_{ji}\left|
\begin{array}{cc}
j & i  \\
\bar{i} &\bar{j}
\end{array}
\right>\equiv\sigma_{ij}\Ket{ij}+\sigma_{ji}\Ket{ji}
\end{equation}
In that case we can provide an explicit expression for the exponent of the
evolution operator,
\begin{equation}
{\mathbf\Omega}=
\left(\begin{array}{cc}
    \frac{-3}{2}(\Omega_{23}+\Omega_{14}) & \frac{1}{2}(\Omega_{12}-\Omega_{23}-\Omega_{14}+\Omega_{34}) \\
    \frac{1}{2}(\Omega_{12}-\Omega_{13}-\Omega_{24}+\Omega_{34}) &  \frac{-3}{2}(\Omega_{13}+\Omega_{24})
\end{array}\right) \ ,
\end{equation}
where $\Omega_{ij}=\frac{\alpha_s}{2\pi}
\left(\frac{1}{2}\ln^2\frac{M_{ij}^2}{\mu^2} - i\pi
\ln\frac{M_{ij}^2}{\mu^2}\right)$.  However, as will be discussed at the end
of Section~\ref{sec:Parameters} the numerical investigation showed that the
Coulomb term in $\Omega_{ij}$ has negligible effect on the massless cluster
evolution, which we do not consider in this section for simplicity. We obtain
the compact form for $\Omega_{ij}$ in terms of the cluster masses or the
partons four-momenta:
\begin{equation}
 \Omega_{ij}= \frac{\alpha_s}{2\pi}\Big[\ln^2{\frac{M_{ij}^2}{\mu^2}}\Big]\sim\ln^2{{M_{ij}^2}}\sim\ln^2{2p_i\cdot p_j}
\end{equation}
The evolution is governed by the exponential of the matrix $\mathbf{\Omega}$
which has the structure
\begin{multline}
\mathbf{U}=e^{\mathbf{\Omega}}=
\frac{e^{-\frac{3}{2}(a+b)}}{\sqrt{\Delta}}\sinh \left(\frac{\sqrt{\Delta}}{2}\right)\\
\left(
\begin{array}{cc}
 \sqrt{\Delta} \coth \left(\frac{\sqrt{\Delta}}{2}\right)+3 (b-a)  & 2 (c-a)  \\
 2 (c-b)  & \sqrt{\Delta} \coth \left(\frac{\sqrt{\Delta}}{2}\right)+3 (a-b)  \\
\end{array}
\right),
\end{multline}
where we introduced new variables:
$a=\ln^2 \Omega_{23}^2+\ln^2 \Omega_{14}^2$, 
$b=\ln^2 \Omega_{13}^2+\ln^2 \Omega_{24}^2$, 
$c=\ln^2 \Omega_{12}^2+\ln^2 \Omega_{34}^2$ and 
$\Delta=9 a^2-4 c (a+b)-14 a b+9 b^2+4 c^2$.
Let us now determine a reconnection probability when the initial colour flow is 
$\Ket{ij}$:
\begin{equation}
  p_{rec} = 
  \frac{|\langle ji|\sigma \rangle|^2}{|\langle ji |\sigma \rangle |^2 + 
   | \left <ij|\sigma \rangle \right |^2} \ .
\end{equation}
We remind the reader that we work in the basis in which the scalar products
are non-orthogonal states and therefore we have $\Braket{ij|ij} = N^2$ and
$\Braket{ij|ji} = N$. Hence $p_{rec}$ can be expressed in terms of evolution
matrix $\mathbf{U}$ in the following way:
\begin{equation} 
p_{rec} = \frac{| U_{11}+NU_{21}|^2}{| NU_{11}+U_{21} |^2 + |  U_{11}+NU_{21}
  |^2} \ ,
\label{2cluster:prec}
\end{equation}
where $U_{ij}$ are matrix elements of $\mathbf{U}$. We will now explore
properties of $p_{rec}$.  The minimal and maximal mixing is obtained at
$U_{21} = 0$ (with $p_{rec}=\frac{1}{10}$) or $U_{11} = 0$
($p_{rec}=\frac{9}{10}$) respectively, where most of the reconnection values
are encountered for extremal cluster mass configurations:  In the case when
$U_{21} = 0$ the condition can be translated to the kinematical situation when
masses of final cluster $M_{13}$ and $M_{24}$ fulfil the following equality:
\begin{equation}
\label{eq:U21}
U_{21} = 0 \Leftrightarrow (\ln^2 M_{12}^2+\ln^2 M_{34}^2 = \ln^2
M_{13}^2+\ln^2 M_{24}^2) \ .
\end{equation}
On the other hand in the case $U_{11} = 0$ the value of $p_{rec}=\frac{9}{10}$
is obtained when
$$U_{11} = 0 = \sqrt{\Delta} \coth \left(\frac{\sqrt{\Delta}}{2}\right)+3
(b-a)\ .$$ In the limit when $\sqrt{\Delta}$ is large\footnote{Already for
  $\Delta=40$ the value of $\coth \left(\frac{\sqrt{\Delta}}{2}\right) =
  1.00359$ and for $\Delta=100$ it is equal to $1.00009.$} $\coth
\left(\frac{\sqrt{\Delta}}{2}\right)\rightarrow 1$ the condition above is much
simpler:
$$\frac{3(a-b)}{\sqrt{\Delta}}=0 \Leftrightarrow (a-c)(c-b)=0\ .$$ Since we
assume that both $U_{11}$ and $U_{21}$ are not equal to $0$ at the same time,
the condition above is only fulfilled when
$$a-c=0 \Leftrightarrow \ln^2 M_{23}^2+\ln^2 M_{14}^2 = \ln^2 M_{12}^2+\ln^2
M_{34}^2\ ,$$ which is also consistent with the numerical results presented in
Fig.~\ref{fig:pRecoParam}.  In general the reconnection probability is bigger than
the probability that the system does not change when: $$p_{rec} > p_{no-rec}
\Leftrightarrow U_{21}^2>U_{11}^2\ ,$$
$$|\sqrt{\Delta} \coth \left(\frac{\sqrt{\Delta}}{2}\right)+3 (b-a)|^2 < |2
(c-a)|^2\ .$$ Finally, let us cast some light on the rapidity dependence of
the result. In order to do it we will work in the frame when particles with
four-momenta $p_1$ and $p_2$ are back-to-back, {\it i.e.}
$p_1=\frac{1}{2}(M_{14},0,0,M_{14})$ and
$p_4=\frac{1}{2}(M_{14},0,0,-M_{14})$, then for $i=2,3$ we express the
four-momenta in the following way: $p_i =p_{Ti}(\cosh{y_i}, \sin{\phi_i},
\cos{\phi_i},\sinh{y_i})$.  Then the scalar products have the simple form:
$$p_1\cdot p_4=\frac{1}{2}M_{14}^2,$$ 
$$p_1\cdot p_i=\frac{1}{2}M_{14}p_{Ti}e^{-yi},$$ 
$$p_4\cdot p_i=\frac{1}{2}M_{14}p_{Ti}e^{yi},$$
$$p_2\cdot p_3=2p_{T2}p_{T3}(\cosh{\Delta y_{23}}-\cos{\Delta\phi_{23}}).$$
and the condition for the minimal mixing from Eq.~\ref{eq:U21} can be rewritten 
as
\begin{equation}
\label{Eq:y2clusers}
\ln{\frac14 M_{14}^2 p_{T2}^2\ln{e^{-2y_2}}}=\ln{\frac14 M_{14}^2
  p_{T3}^2\ln{e^{-2y_3}}} \ ,
\end{equation}
such that, when $p_{T2}\sim p_{T3}$ and $\Delta Y=y_3-y_2\sim0$, meaning that
the initial partons in clusters have similar transverse momenta and are close
in the rapidity, the reconnection mixing is minimal. Therefore, more likely
will be reconnections when the $\Delta Y$ values for the quark anti-quark pairs
of the original clusters are big, which is also confirmed numerically
Fig.~\ref{fig:deltaYRambo}. It is also interesting to see the transverse
momentum dependence of the result from Eq.~\ref{Eq:y2clusers} which we plan to
investigate in the future while constructing a phenomenological model based on
the current studies.

\section{Numerical Results}
\label{sec:NumericalResults}

A significant difference towards other colour reconnection
models\,\cite{Gieseke:2012ft,Gieseke:2017clv} is that we do not
directly compare clusters and then choose a configuration that would
leave us with pre-specified properties such as a lower invariant
cluster mass.  Since we calculate the probabilities to evolve into
different colour flows we first show that our approach leads to
reasonable results compatible with the effects of conventional colour
reconnection algorithms.  In order to analyse the effect of the colour
reconnection we mostly compare kinematic variables associated to the
clusters before and after reconnection. We first consider `mesonic`
reconnections and later proceed to include `baryonic' reconnections as
outlined in Sec.~\ref{sec:Baryons}.  We generate initial cluster
configurations using the RAMBO method\,\cite{Kleiss:1985gy}, and a
variation of the Jadach algorithm\,\cite{JADACH1975297}, which was
used for the UA5 model\,\cite{ALNER1987445}. While RAMBO is performing
a flat phase space population, including a cluster configuration which
would not be expected from a pre-confining shower evolution, the UA5
algorithm provides us already with a very physical mass spectrum.

\subsection{Mesonic reconnections}
\label{sec:Mesonic}

\begin{figure}[t]
\centering
\includegraphics[width=7cm]{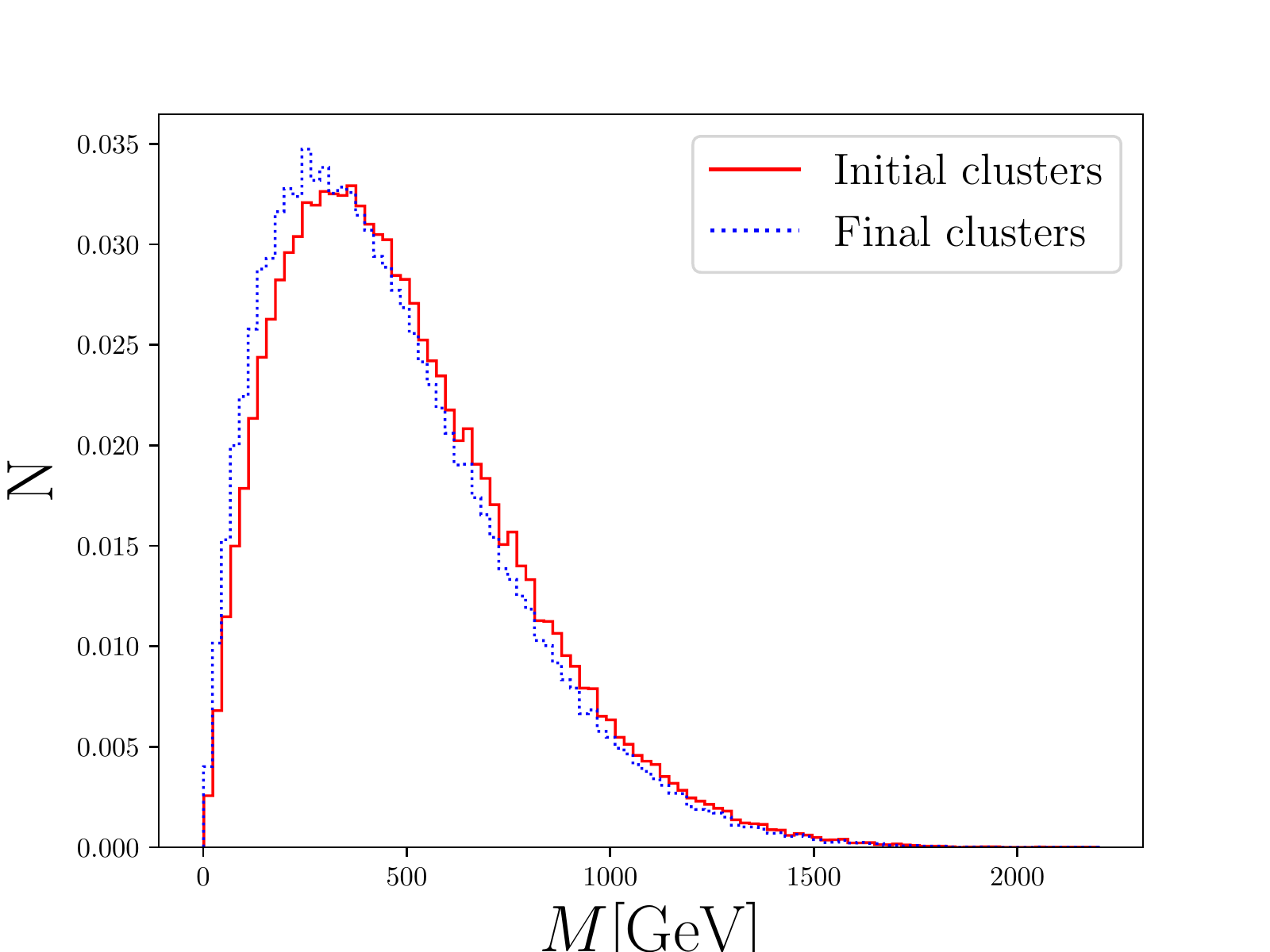}
\includegraphics[width=7cm]{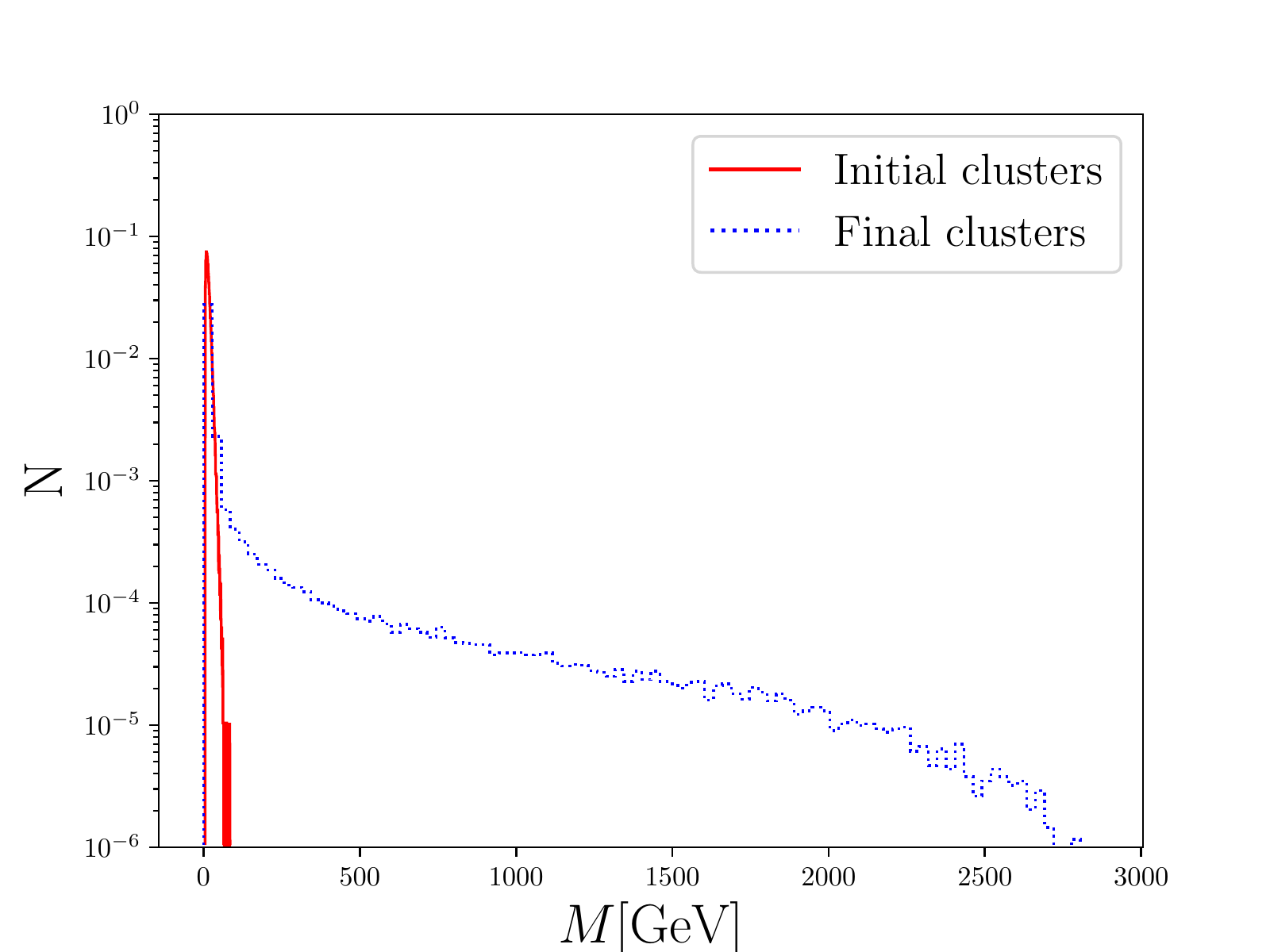}
\caption{Invariant mass distribution of 5 clusters before and after mesonic colour 
reconnection for RAMBO(left) and UA5(right) kinematics. 
}
\label{fig:clustermasses}
\end{figure}

\begin{figure}[t]
\centering
\includegraphics[width=5cm]{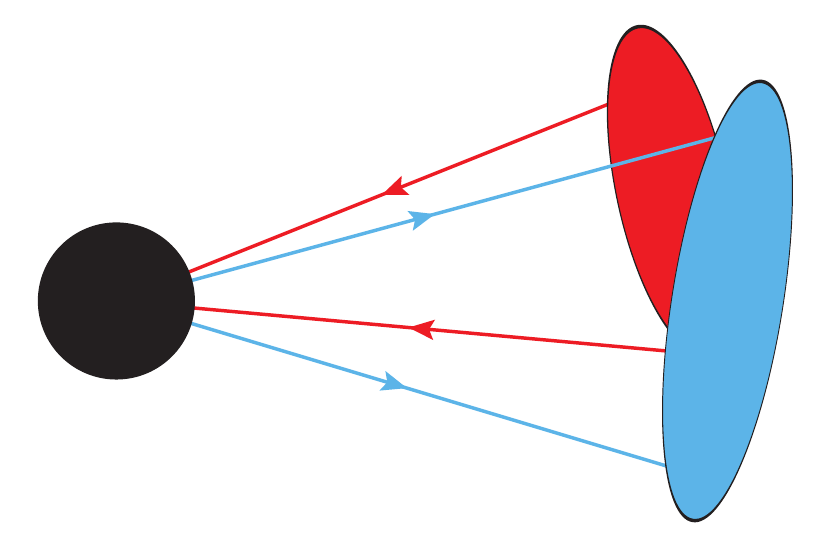}\hspace*{2cm}
\includegraphics[width=5cm]{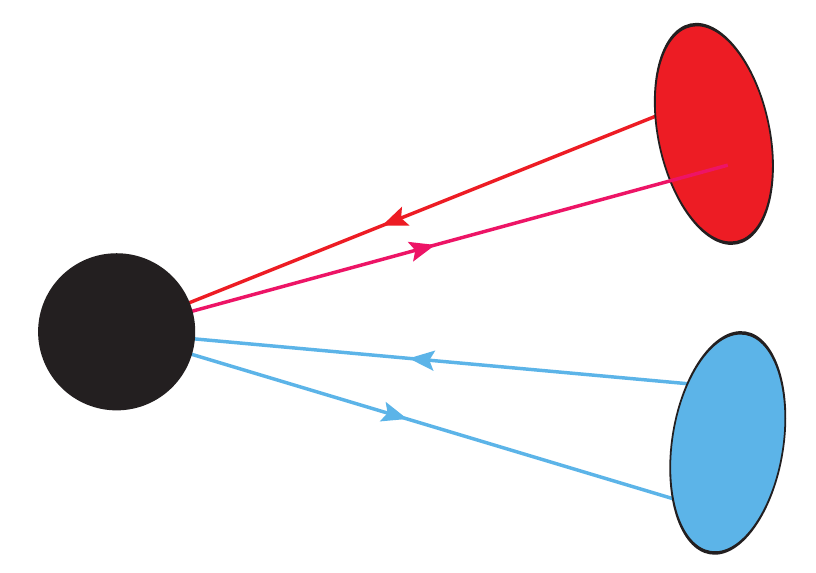}
\caption{ Sketch of the cluster configuration resulting from two alternative 
colour flows from two cluster evolution. 
The left figure shows two large overlapping clusters and the right 
figure shows a different colour flow resulting in smaller clusters consisting of 
quark anti-quark pairs closer in momentum space. The black blob can be associated
with an interaction or parts of an interaction that would lead to the shown configuration. 
}
\label{fig:subsystem}
\end{figure}

\begin{figure}[t]
\centering
\includegraphics[width=7cm]{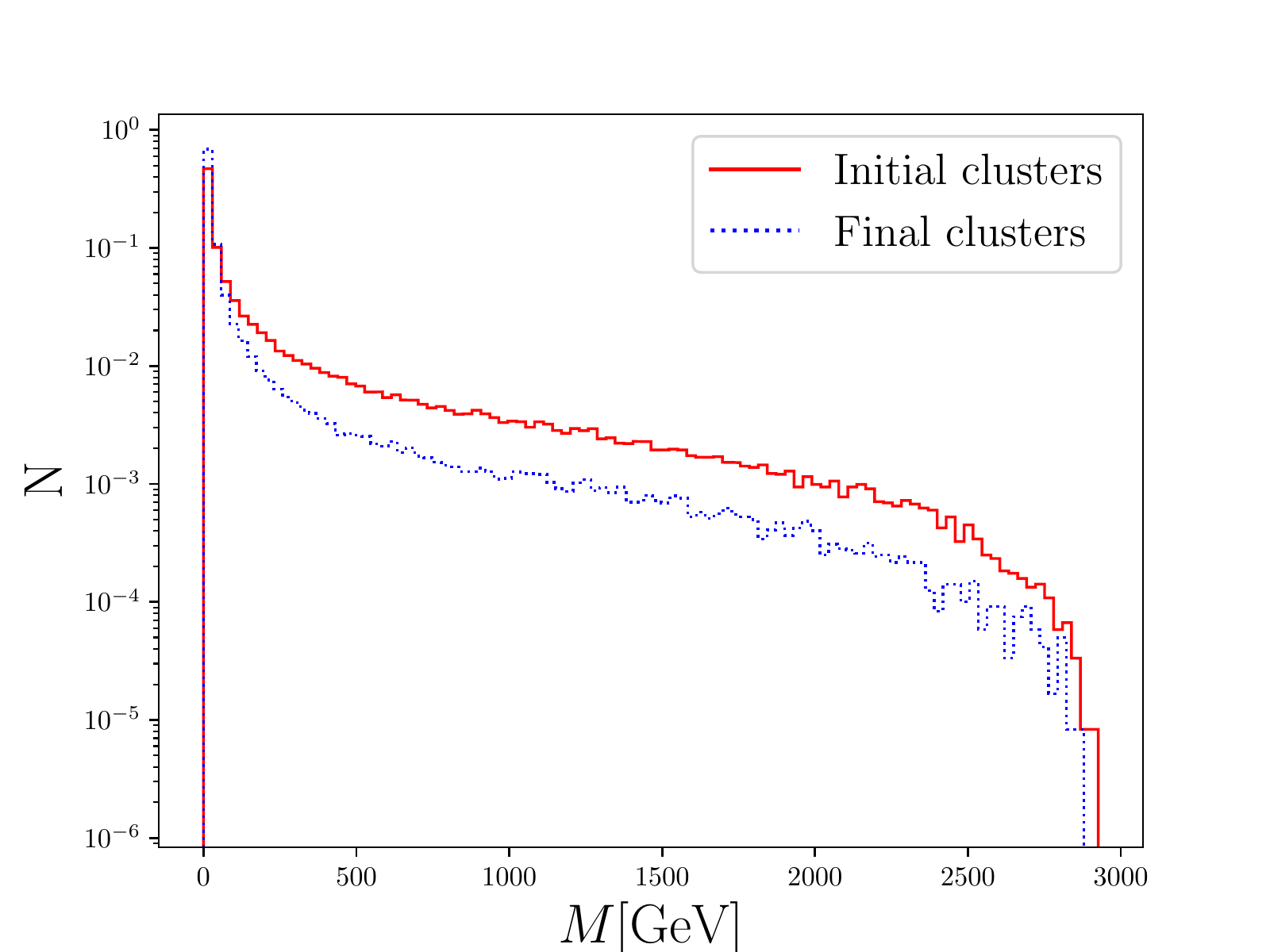}
\caption{ Invariant cluster mass distribution before and after colour 
reconnection for UA5 kinematics with random initial colour connections.
}
\label{fig:clustermasses_UA5}
\end{figure}

\begin{figure}[t]
\centering
\includegraphics[width=7cm]{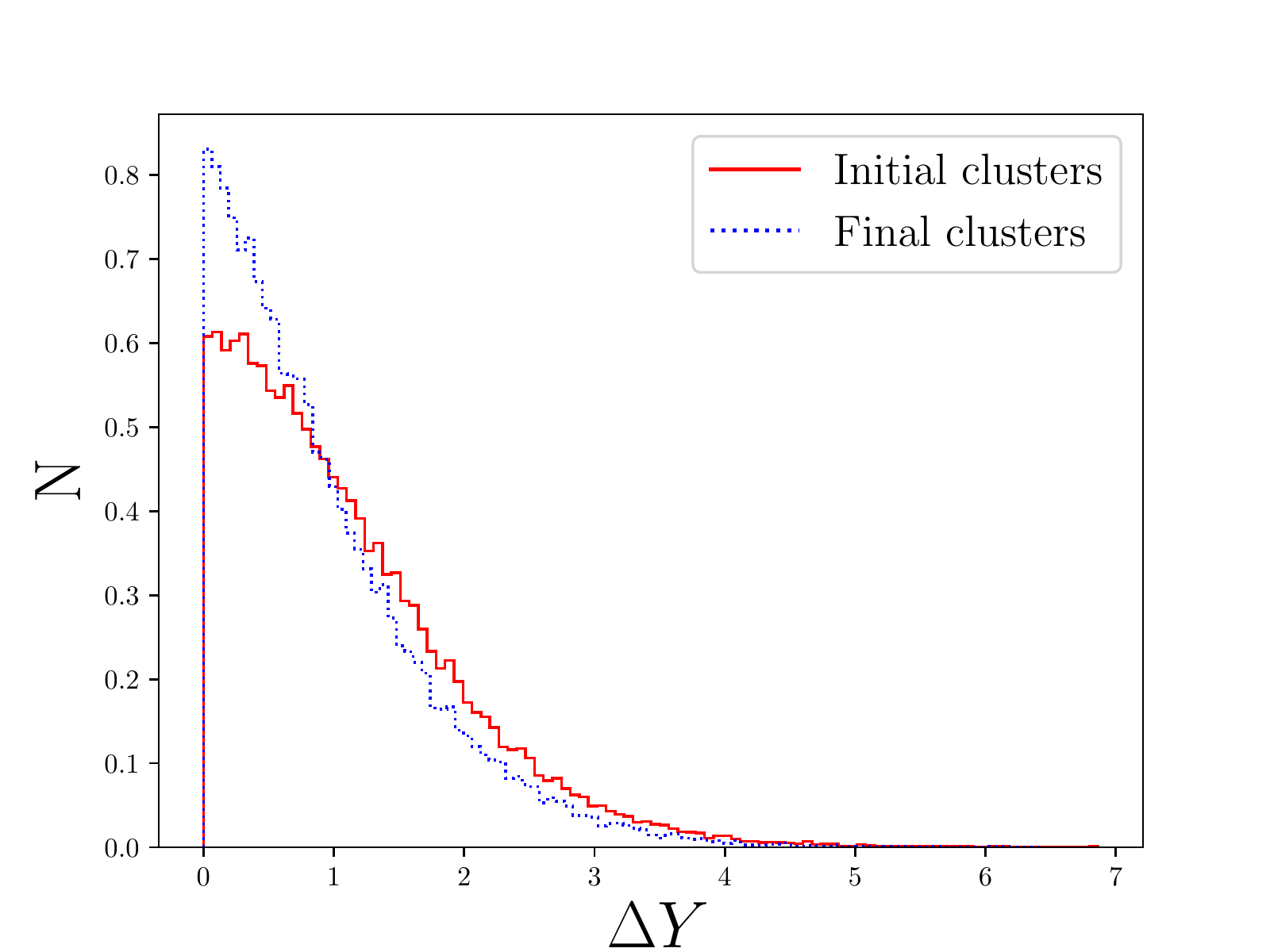}
\includegraphics[width=7cm]{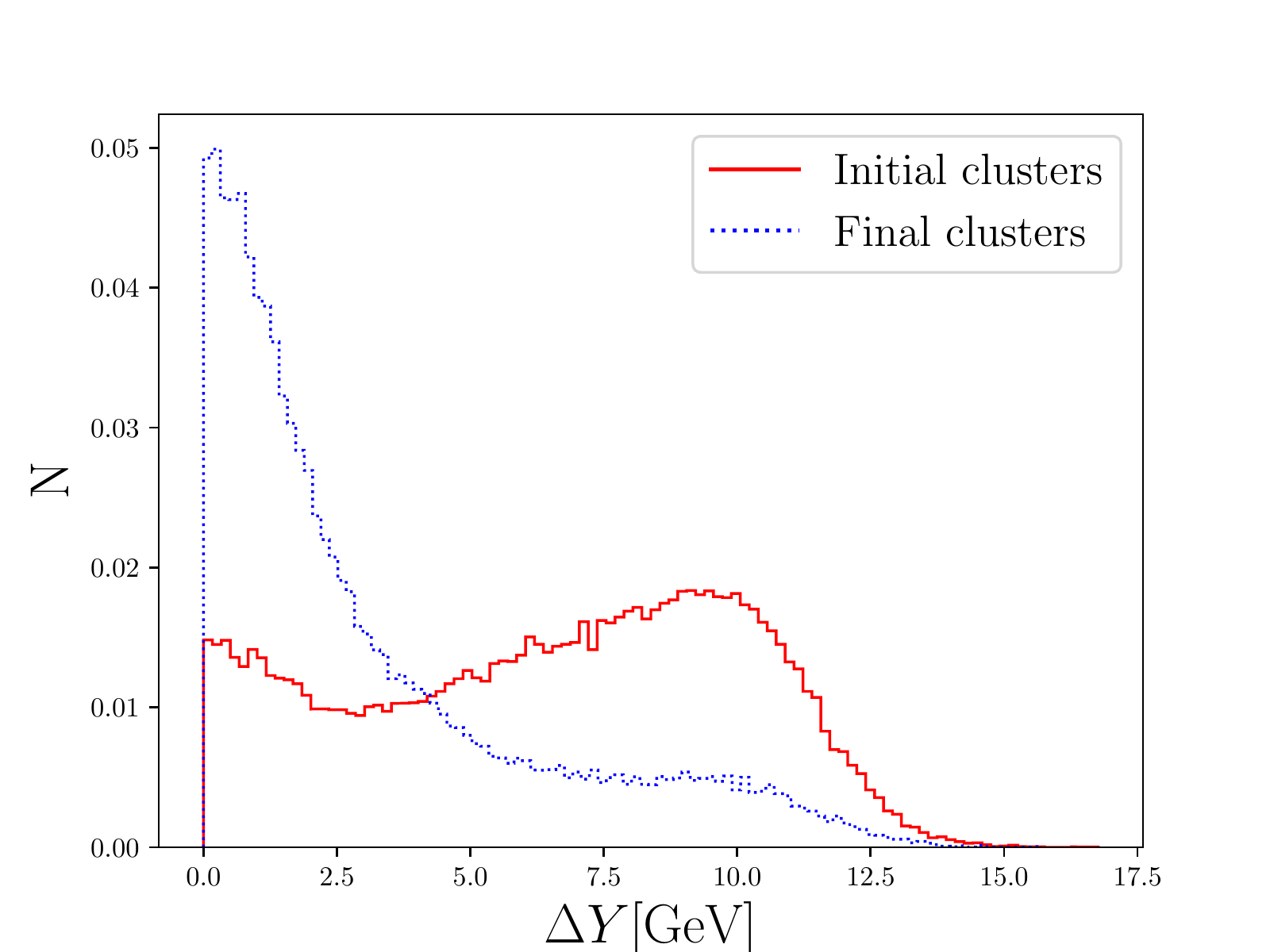}
\caption{ Histogram of the $\Delta Y$ values for the quark 
anti-quark pairs of the original clusters that were reconnected 
and the $\Delta Y$ value of the quark anti-quark pairs of the 
reconnected clusters. (a) RAMBO Phase space. (b) UA5 phase space with
random colour connections.
}
\label{fig:deltaYRambo}
\end{figure}

A known issue which concerns the modelling of LHC events is that it is
a priori not clear how the colour connection between different
scattering centres of multi parton interactions looks like. The
clusters emerging from these interactions are in general too heavy
meaning that they consist of quark anti-quark pairs which are not close
in momentum space, in terms of a small invariant mass of the pairs. In
this case colour reconnection models are used to restore the notion of
a colour pre-confined state leading to a shift towards lower invariant
cluster masses.  In Fig.\,\ref{fig:clustermasses} we show the
invariant mass distribution for five clusters before and after colour
reconnection where the phase space was sampled with the two phase
space algorithms mentioned above, RAMBO and UA5-type.
While for the RAMBO kinematics the invariant mass distribution gets shifted
towards lower values the clusters generated with the UA5 model already consist
of quark anti-quark pairs close in momentum space, which leads to clusters
with the smallest invariant mass possible. Considering different colour
flows will eventually connect quark anti-quark pairs well separated in
rapidity leading to heavier clusters as seen in the right plot of 
Fig.\,\ref{fig:clustermasses}.
When sampling the cluster kinematics with the UA5 model \textit{but} enforcing
random colour connections between the quarks and anti quarks for the initial
configuration the resulting clusters are large and (in rapidity span) overlapping.
A sketch of this configuration is shown in Fig.\,\ref{fig:subsystem} due 
to comprehensibility for the simple case of two cluster evolution.
For this configuration our \textit{Ansatz} for colour reconnection again
chooses colour flows leading to a shift towards lower values in terms of 
invariant cluster mass. 
The effect on the invariant mass
spectrum can clearly be seen in Fig.\,\ref{fig:clustermasses_UA5},
where we plotted the logarithm of the invariant cluster masses.
We conclude that the approach followed in this paper naturally prefers colour 
flows leading to a configuration with lower invariant cluster masses.
We also stress here that the algorithm does not veto any colour flows 
which would lead to an increase in terms of invariant cluster mass.
To get an intuitive picture of what happens on the quark level the 
rapidity difference, $\Delta Y$, between the quark anti-quark pairs 
which were participating in the reconnection process is shown in 
Fig.\,\ref{fig:deltaYRambo} for the RAMBO phase space and for the UA5 model 
with random initial colour connections.
In both figures we see that colour flows resulting in clusters
consisting of quark anti-quark pairs which are closer in rapidity are
clearly preferred.  Again the effect is more pronounced for the UA5
model with random initial colour connections.
 
\subsection{Baryonic reconnections}

\begin{figure}[t]
\centering
\includegraphics[width=9cm]{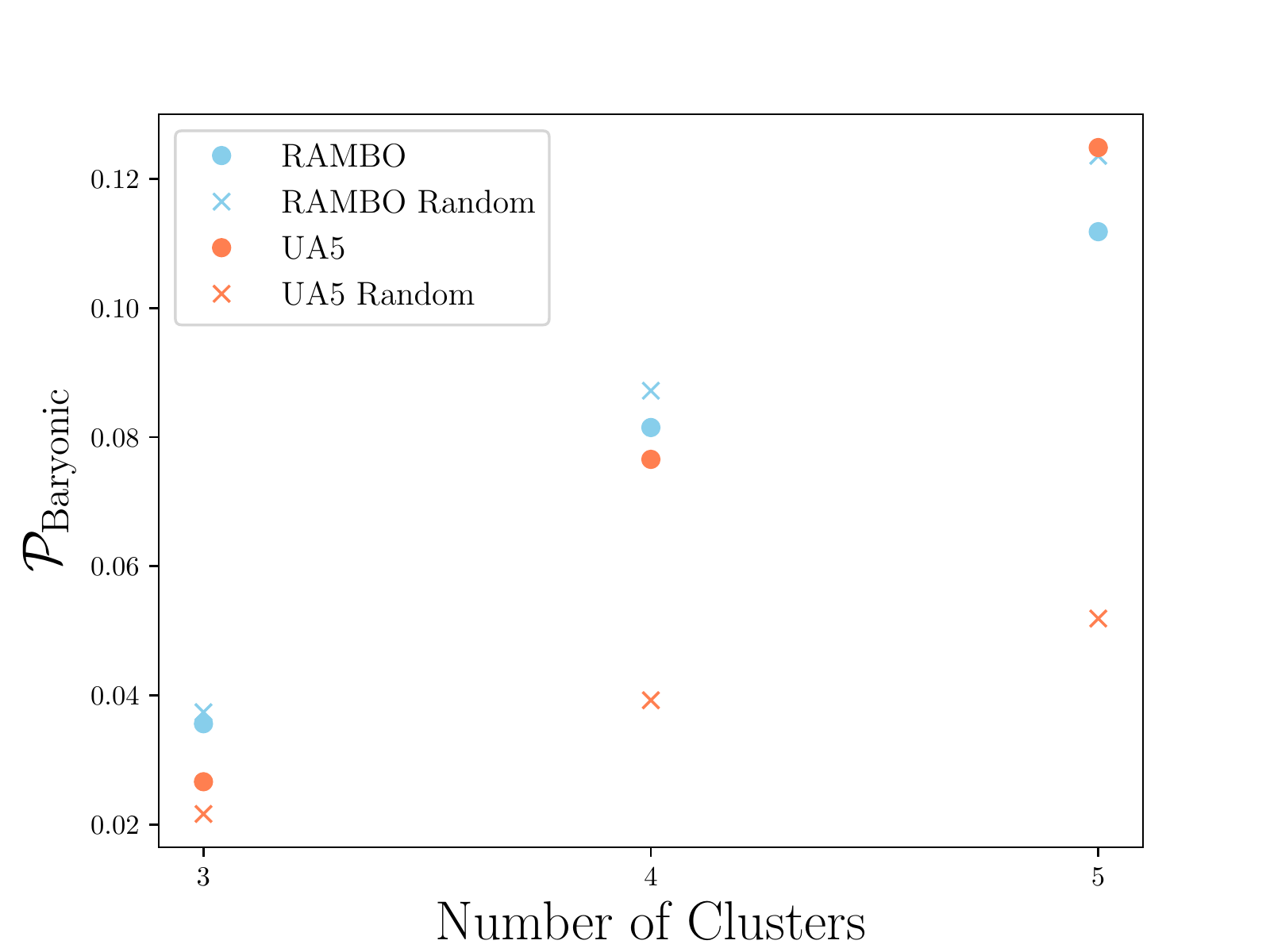}
\caption{The average baryonic reconnection probabilities for the RAMBO and UA5
  phase space option with original and random initial colour
  connections.  }
\label{fig:pBaryonic}
\end{figure}

\begin{figure}[t]
\centering
\includegraphics[width=9cm]{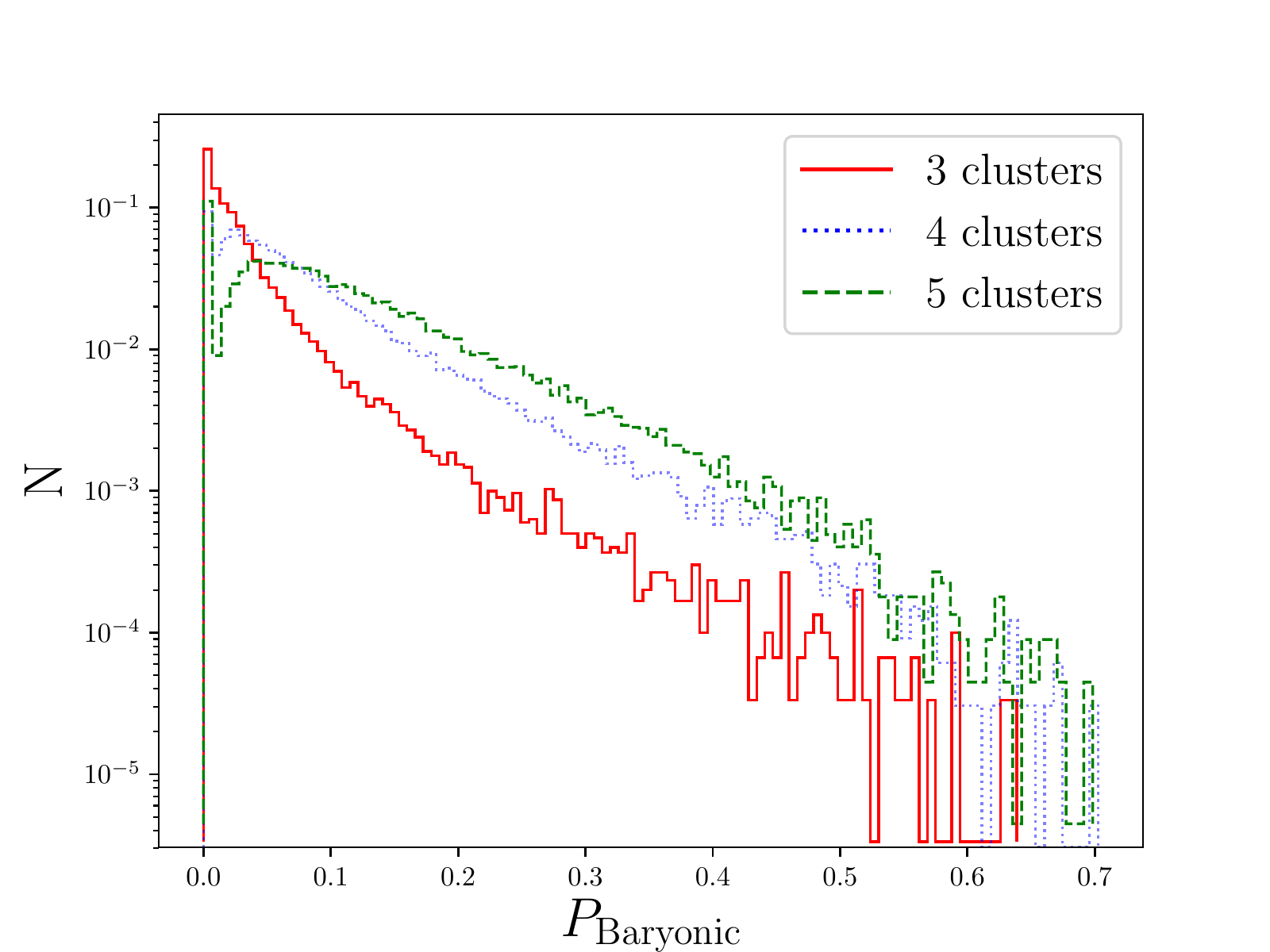}
\caption{Histogram of baryonic reconnection probabilities for three,
  four and five cluster evolution where the quark kinematics was
  sampled with the RAMBO phase space algorithm.}
\label{fig:histpbaryonic}
\end{figure}

Within the context of our model a baryonic colour flow can be
introduced as explained in Sec.\,\ref{sec:Baryons}.  In
Fig.\,\ref{fig:pBaryonic} the average baryonic reconnection
probability for both phase space algorithms is shown.
Depending on the phase space algorithm we employ to sample the initial
configurations, the average baryonic reconnection probability ranges
between $2\%$ and $12\%$. The first striking observation is that the
probability rises with the number of clusters considered. The more
clusters in an event, the more likely it is to find a candidate for
baryonic reconnection.  For RAMBO kinematics the initial colour
configuration has no effect on the average reconnection probability.
For the UA5 phase space it strongly depends on the initial colour
configuration. Since the original UA5 cluster configuration already is
in a state where the quarks are colour connected to their closest
neighbours in phase space the probability for reconnection into a
different mesonic state is suppressed, which raises the probability to
end up in a baryonic state.  If the quarks are randomly connected the
average baryonic reconnection probability is suppressed since the
probability for mesonic reconnection is high.  Also the more clusters
we consider, the higher the probability is to find a candidate for
baryonic reconnection.  Now we proceed to study the three, four and
five cluster evolution with the RAMBO phase space in detail.  The
distribution of the reconnection probabilities is shown in
Fig.\,\ref{fig:histpbaryonic}.
The baryonic reconnection probability tends to prefer lower values
with a pronounced peak at zero. The tail towards higher values in the
distribution might indicate some preferred kinematic configurations
for the evolution into a baryonic state. With only one possible
baryonic configuration, the three cluster evolution is convenient to
analyse and to extract a kinematic dependence.

\begin{figure}[t]
\centering
\includegraphics[width=9cm]{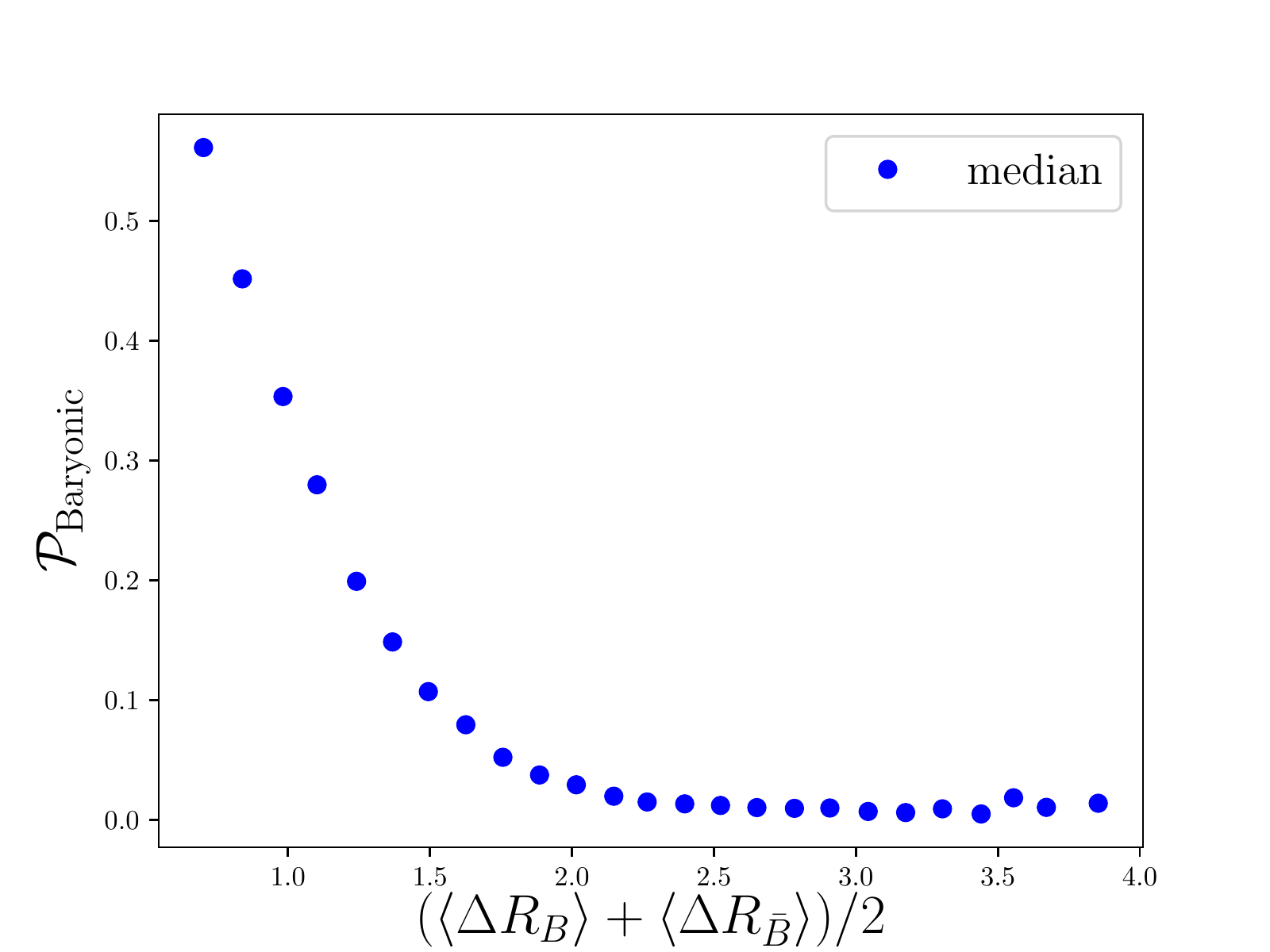}
\caption{Median values for the baryonic reconnection probability 
with respect to $ (\langle \Delta R_B \rangle + \langle \Delta R_{\bar{B}}\rangle ) / 2$
of the baryonic clusters for 3 cluster evolution.
}
\label{fig:scatter}
\end{figure}

In Fig.\,\ref{fig:scatter} the probability to evolve into 
a baryonic state with respect to the sum of average $\Delta R$ values 
of the quarks and anti-quarks that would constitute a baryonic 
cluster, $ ( \langle \Delta R_B \rangle + \langle \Delta R_{\bar{B}}\rangle ) / 2 $ 
is shown, where $\Delta R$ is defined as the distance between the 
constituent quarks in the $y-\phi$ plane
\begin{equation}
\Delta R = \sqrt{(\Delta \phi)^2 + (\Delta y)^2}.
\end{equation}
and we define $\langle \Delta R_{B,\bar{B}}\rangle$ as
\begin{equation}
\langle \Delta R_{B,\bar{B}}\rangle = (\Delta R_{12,\bar{1}\bar{2}} + \Delta R_{13,\bar{1}\bar{3}} + \Delta R_{23,\bar{2}\bar{3}})/3 \ ,
\end{equation}
where the subscripts $(1,2,3)$, $(\bar{1},\bar{2},\bar{3})$, 
denote the quarks(anti-quarks) inside the baryonic(anti-baryonic) clusters.
The median shows a rising probability with lower $( \langle \Delta R_B
\rangle + \langle \Delta R_{\bar{B}}\rangle ) / 2$ values which
indicates that the formation into a baryonic cluster is preferred if
the three quarks {\it and} the three anti-quarks are close together in
$\Delta R$ space. We note that the baryonic and the anti-baryonic
cluster can still be overlapping since we do not take the distance
between them into account.

\subsection{Unbaryonization}
\begin{figure}[t]
\centering
\includegraphics[width=7cm]{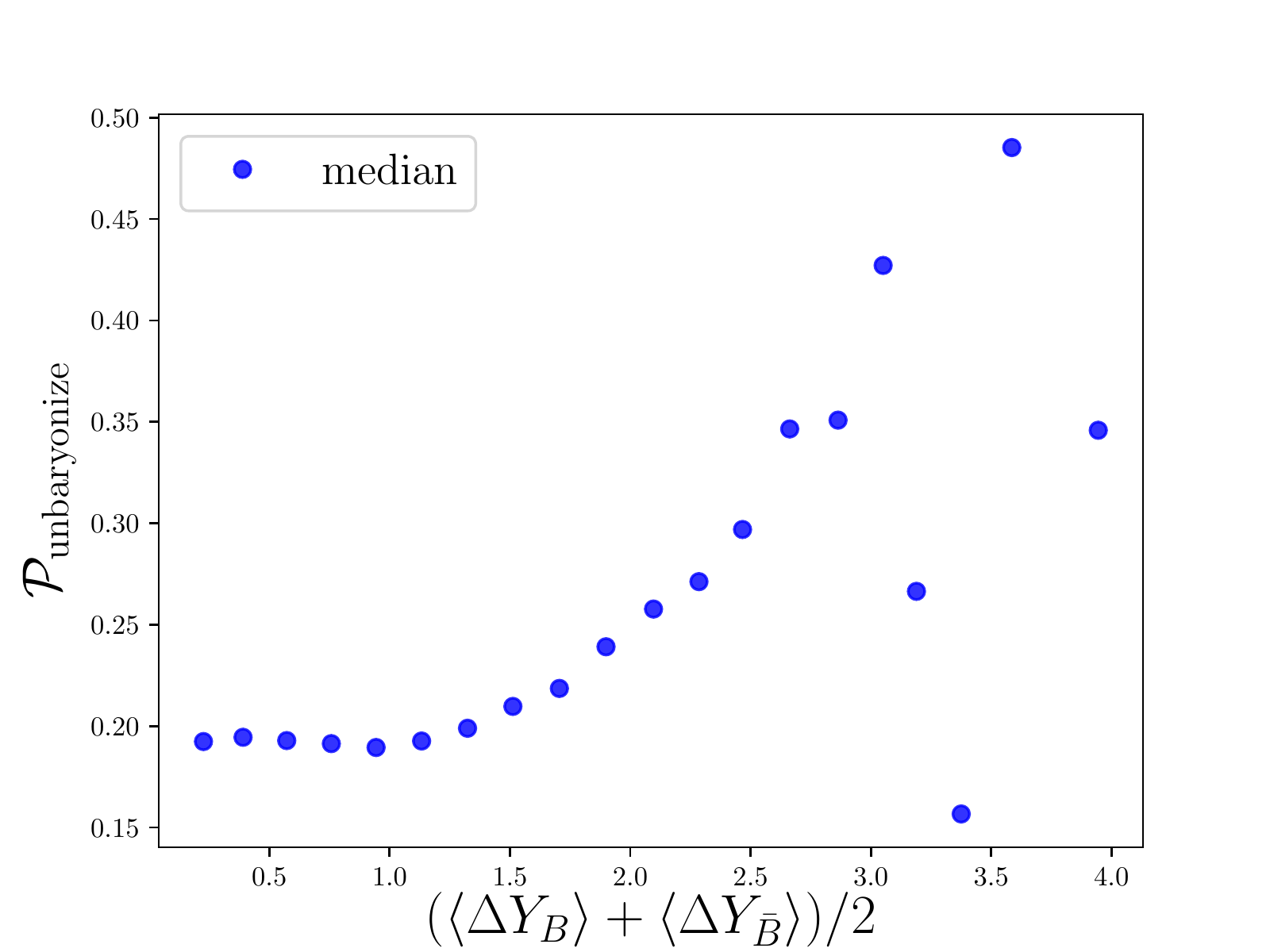}
\includegraphics[width=7cm]{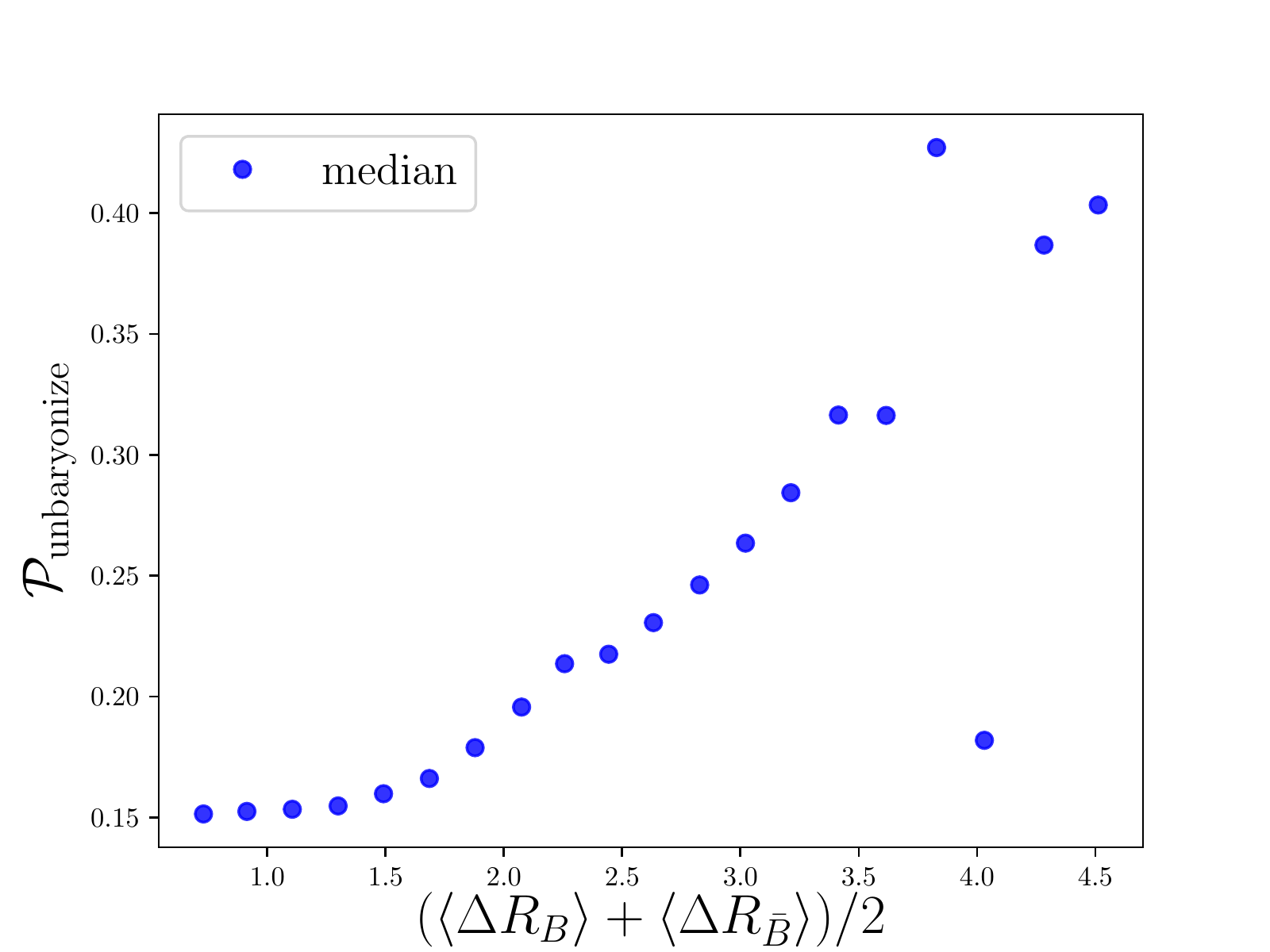}
\caption{Average unbaryonization probabilities with respect to the average 
sum of $\langle \Delta Y \rangle$ and $\langle \Delta R \rangle$ of the 
constituent quarks of the baryonic clusters. 
The outliers for high values are due to missing statistics.
}
\label{fig:scatter2}
\end{figure}

Colour reconnection algorithms that allow reconnection into a baryonic
state are structured in a way that once a baryonic cluster is formed,
it is not considered for any further modifications. This clearly
biases the reconnection procedure but has been necessary in order to
cope with the rising complexity of many cluster systems. In principle
a system could evolve into a baryonic state and then evolve again into
a mesonic state which in turn lowers the amount of baryonic clusters
occurring in an event. In the context of our model we can study the
evolution back into a mesonic state by considering
\textit{unbaryonization} where we start the evolution with a baryonic
configuration as the initial state and calculate the probability to
evolve into a mesonic cluster configuration. In
Fig.\,\ref{fig:scatter2} we show the average probabilities for
unbaryonization in terms of $(\langle \Delta Y_B \rangle + \langle
\Delta Y_{\bar{B}} \rangle)/2$ and $(\langle \Delta R_B \rangle +
\rangle \Delta R_{\bar{B}} \rangle)/2$, where $\langle \Delta Y_B
\rangle = (|y_{q1}-y_{q2}|+|y_{q1}-y_{q3}|+|y_{q2}-y_{q3}|)/3$. This
further adds to the intuitive picture that the probability for forming
a baryonic cluster is high if the three quarks and the three
anti-quarks are close in momentum space. This also suggests that the
colour field between quarks is enhanced if they do fly in the same
direction.
In principle this could be used in a model such as\,\cite{Gieseke:2017clv}
to decide whether a baryonic cluster should be kept or not which 
allows for more flexibility and may un-bias reconnection algorithms.

\subsection{Parameter variations and general findings}
\label{sec:Parameters}

\begin{figure}[t]
\centering
\includegraphics[width=7cm]{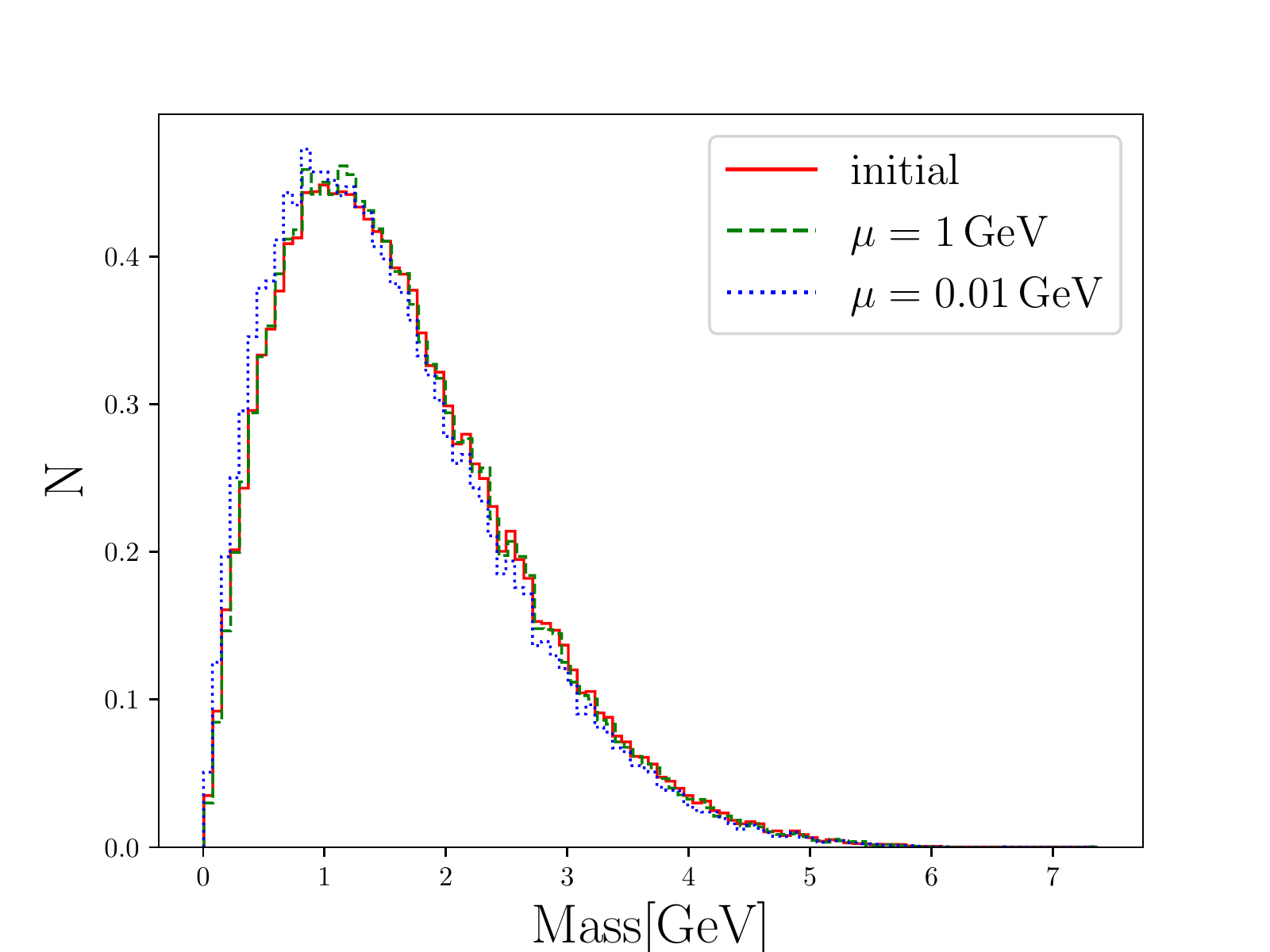}
\includegraphics[width=7cm]{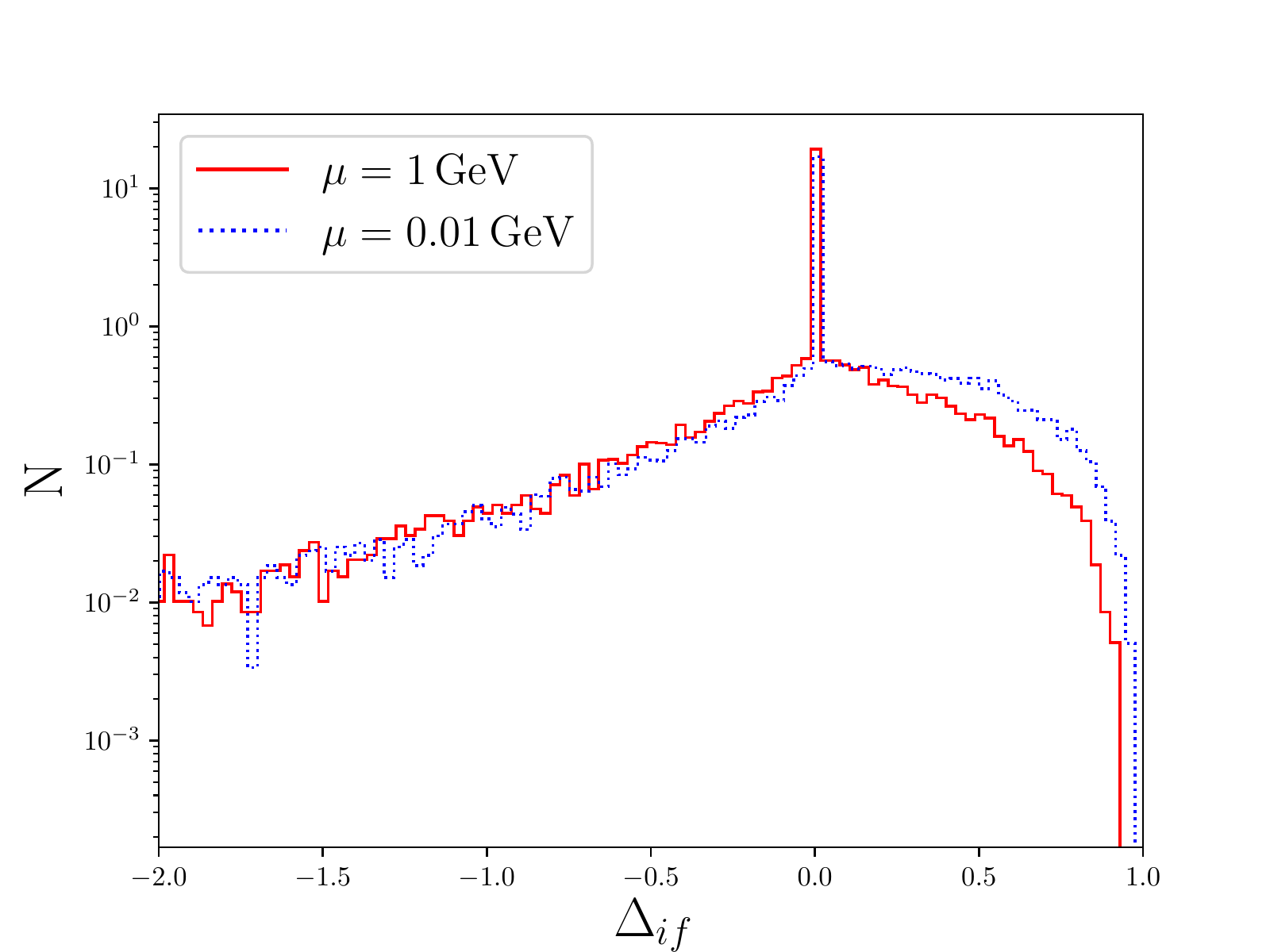}
\caption{ Distribution of invariant cluster masses before and after
  colour reconnection and colour length drop for different cut-off
  values of $\mu=\{1,0.01\}\,\mathrm{GeV}$.  }
\label{fig:mass}
\end{figure}

The Ansatz for the colour flow evolution, Eq.\,\ref{eq:EvolAnsatz}
depends on the two parameters $\mu$ and $\alpha_s$. Also the so-called
Coulomb term proportional to $i \pi$ affects the weights of the
different colour flows.  The reconnection probability is a dynamic
quantity as it strongly depends on the kinematics of the cluster
constituents before and after reconnection and the parameter $\mu$
which can be viewed as a cutoff parameter of the colour flow evolution
in Eq.\,\ref{eq:leadingEvolution}.  In Fig.\,\ref{fig:mass} we show
the distribution of invariant cluster masses for four cluster
evolution with two different values of
$\mu=\{1, 0.01\}\,\mathrm{GeV}$ and the corresponding colour
length drop \cite{Gieseke:2012ft} which is defined as
\begin{equation}
\Delta_{\mathrm{if}} = 1 - \frac{\lambda_{\mathrm{final}}}{\lambda_{\mathrm{initial}}},
\end{equation} 
where $\lambda_{\mathrm{initial}}$ and $\lambda_{\mathrm{final}}$
denote the colour length before and after colour reconnection in an 
event which is defined as the sum of squared invariant cluster masses
\begin{equation}
\lambda = \sum_{i=1}^{N_{\mathrm{cl}}} M_i^2.
\end{equation}
If there is no colour reconnection $\lambda_{\mathrm{initial}} \approx
\lambda_{\mathrm{final}}$ and $\Delta_{\mathrm{if}}$ approximately
vanishes.  If $\Delta_{\mathrm{if}}\approx 1$ there was quite a
significant change in $\lambda$ which indicates a big effect due to
colour reconnection.  The kinematics of the four clusters were sampled
with the RAMBO method.  In order to have more physical cluster masses
we sample them with a centre-of-mass energy of $10\,\mathrm{GeV}$
which is closer to the cluster mass spectrum at the end of a typical
shower evolution.

Comparing the four cluster evolution with the different values for $\mu$ we
see that the lower $\mu$, the more likely it is to pick a colour flow which
results in a reduction of invariant cluster masses. For all values the
distribution of $\Delta_{\mathrm{if}}$ peaks at zero and is then distributed
towards the positive and negative region where the majority of the values are
in the positive region indicating a reduction in terms of invariant cluster
masses.  Negative values of $\Delta_{\mathrm{if}}$ are also possible since we
do not veto any colour flows which would result in higher invariant cluster
masses.  For $\mu=0.01$ the colour reconnection algorithm has the highest
impact, severely shifting the distribution of invariant cluster masses towards
smaller values which can also be seen for $\Delta_{\mathrm{if}}$. Because
$\mu$ can be interpreted as the cut-off parameter of the colour flow evolution
if $\mu \rightarrow 0$ the colour flow evolves into a state of preferably low
invariant cluster masses.

\begin{figure}[t]
\centering
\includegraphics[width=7cm]{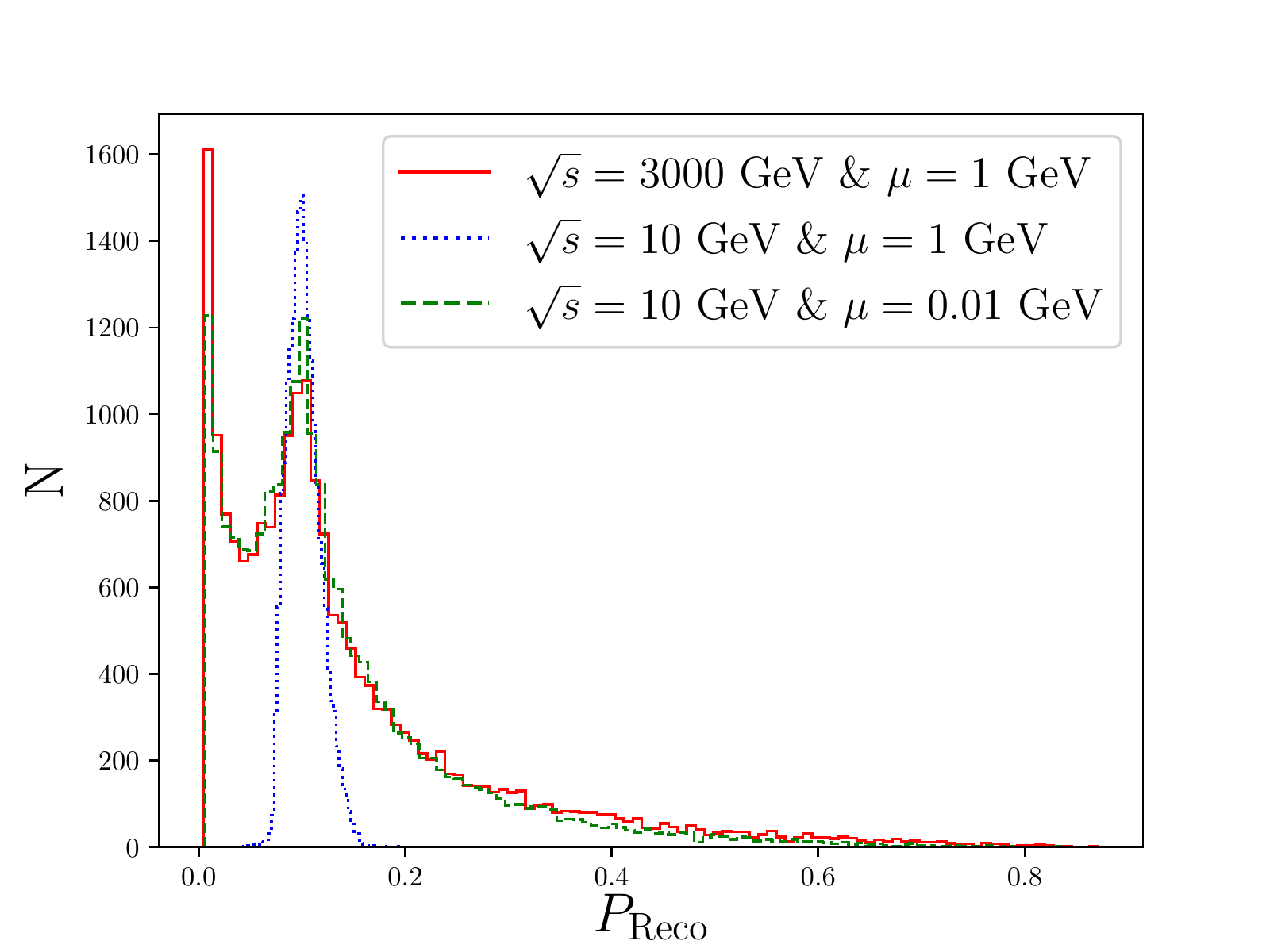}
\caption{ Reconnection probability for the 2 cluster evolution for 
$\sqrt{s}=3000\,\mathrm{GeV}$ and $\sqrt{s}=10\,\mathrm{GeV}$.
}
\label{fig:pReco2}
\end{figure}

In Sec\,\ref{sec:Mesonic} we showed the effect of the algorithm on a
relatively unphysical distribution of cluster masses as a simple proof of
concept, that our Ansatz and our algorithm indeed produce reasonable results.
In this section we study the evolution of colour flow and the behaviour of the
model at different centre-of-mass energies $\sqrt{s}$ with the RAMBO method
where we compare $\sqrt{s}=3000\,\mathrm{GeV}$ with a more physical
centre-of-mass energy of $\sqrt{s}=10\,\mathrm{GeV}$, probing a spectrum of
smaller clusters.  We show the reconnection probability for the case of two
cluster evolution for the two different centre-of-mass energies with different
$\mu$ values in Fig.\,\ref{fig:pReco2}.
While at $\sqrt{s}=3000\,\mathrm{GeV}$ the reconnection probability covers the
whole range from zero to one, at $\sqrt{s}=10\,\mathrm{GeV}$ and
$\mu=1\,\mathrm{GeV}$ they are narrowly distributed around $0.1$.  With $\mu=1
\,\mathrm{GeV}$ for small centre-of-mass energies the values are closer
together which leads to similar reconnection probabilities in each event.
This can be countered by reducing the $\mu$ parameter such that the ratio in
the logarithm is the same as for higher energies which is also shown in
Fig.\,\ref{fig:pReco2} for the two cluster evolution with
$\sqrt{s}=10\,\mathrm{GeV}$ and $\mu=0.01\,\mathrm{GeV}$.

With a smaller $\mu$ parameter the reconnection probabilities start to
cover the whole range, and we observe the same behaviour for baryonic
reconnections.  The smaller the value of $\mu$ the amplitude in colour
space will continue to evolve down to a much smaller scale which, in
the end will result in a colour flow with preferably small invariant
cluster masses. In a full model the parameter $\mu$ could be tuned to
data in order to define the cutoff at which the evolution is bound to
stop and to verify if the amplitude does indeed favour a state of
small invariant cluster masses or a state which allows for more
fluctuations in terms of cluster size.

\begin{figure}[t]
\centering
\includegraphics[width=7cm]{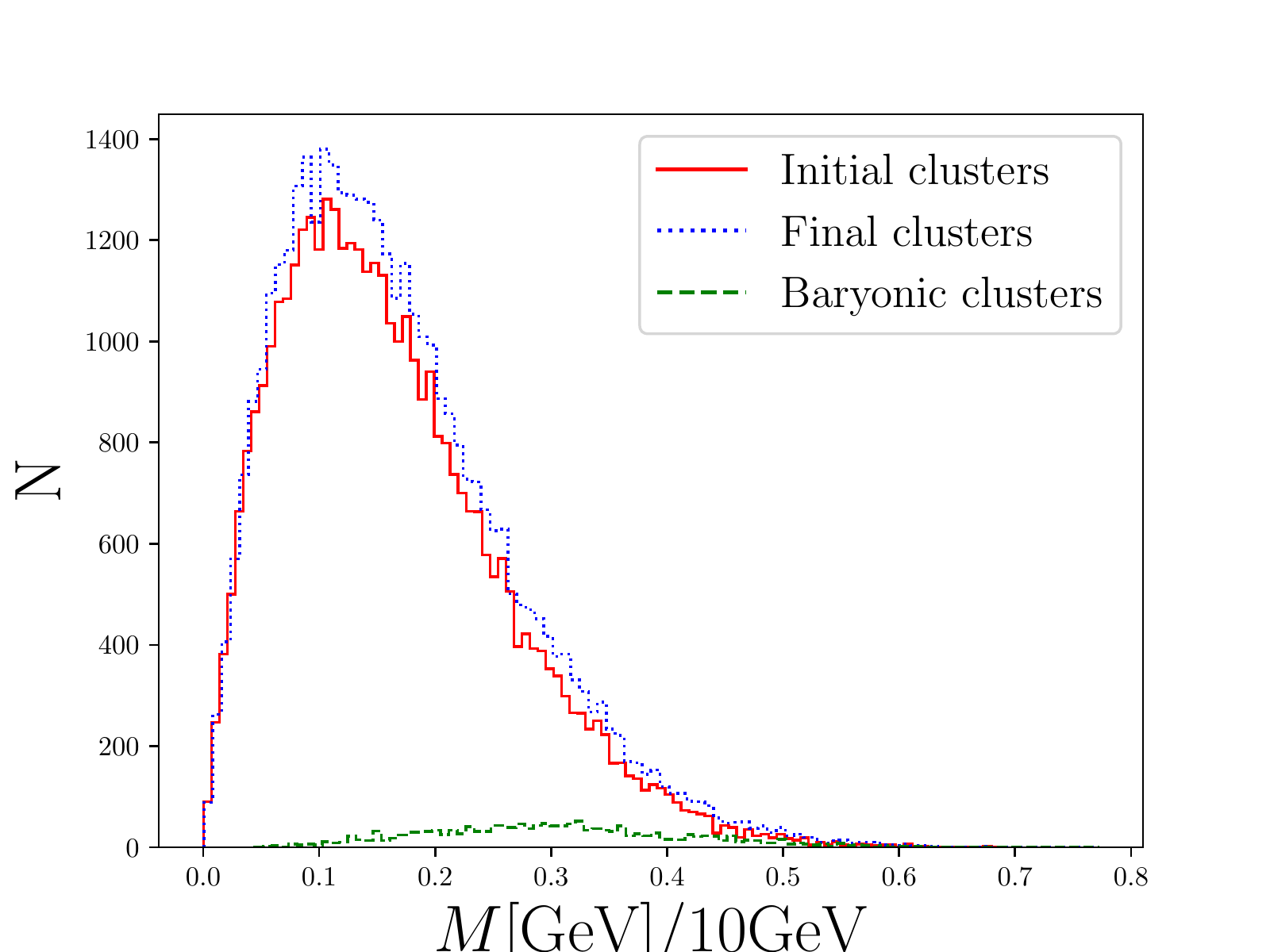}
\includegraphics[width=7cm]{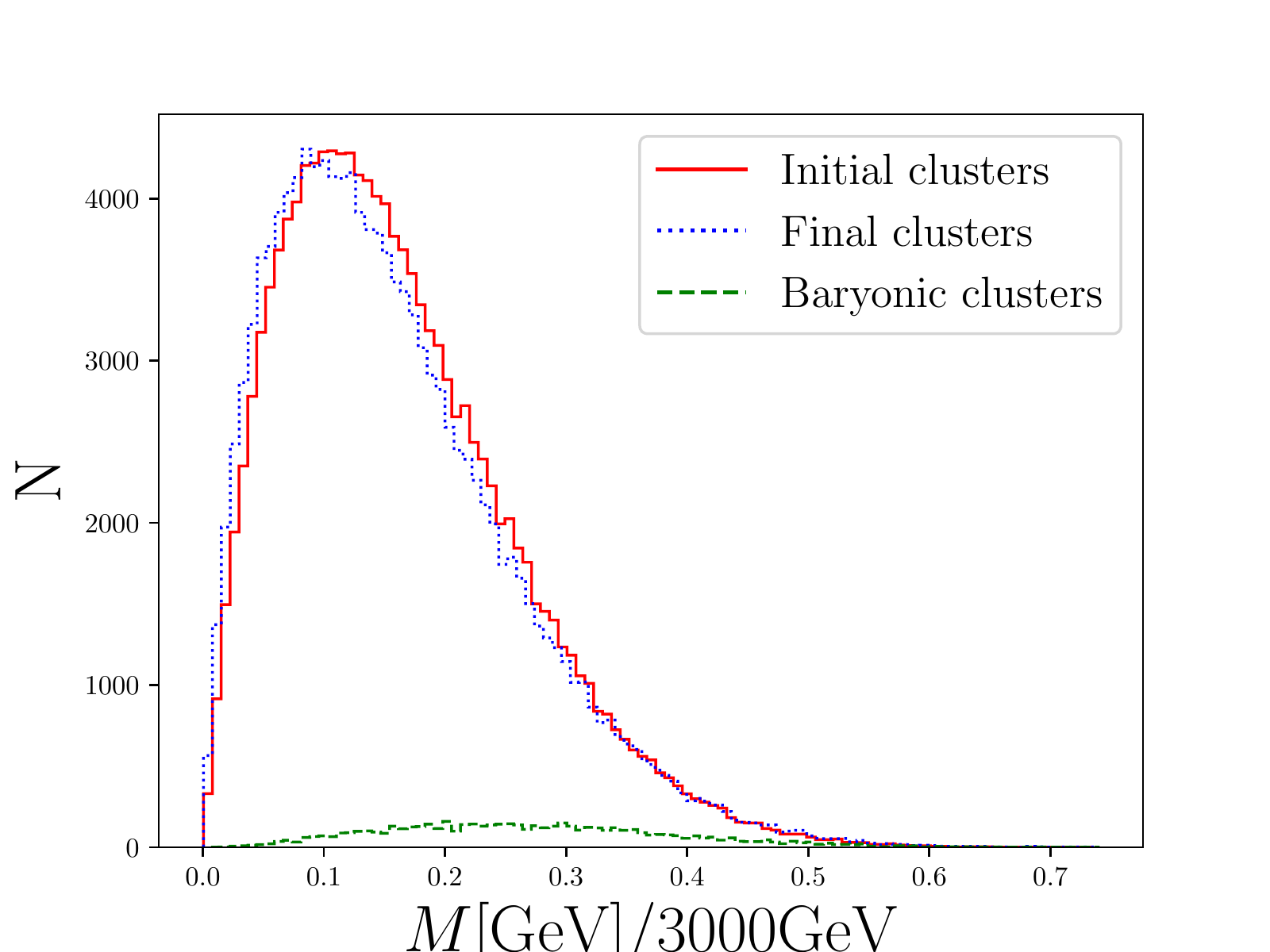}
\caption{
Distribution of invariant cluster masses for $\sqrt{s}=10\,\mathrm{GeV}$ 
and $\sqrt{s}=3000\,\mathrm{GeV}$ divided by their respective $\sqrt{s}$.
}
\label{fig:massdist}
\end{figure}

Another interesting topic is the mass distributions for different
centre of mass energies.  In Fig.\,\ref{fig:massdist} we show the mass
distributions of four clusters for $10\,\mathrm{GeV}$ and
$3000\,\mathrm{GeV}$ divided by the centre-of-mass energy with the
possibility to produce baryonic clusters.
The distributions of the reconnected clusters are roughly the same
although for $10\,\mathrm{GeV}$ there is much less room to evolve into
a state of smaller cluster masses since we used $\mu=1\,\mathrm{GeV}$.
The mass distribution of baryonic clusters is also shifted between the
two centre of mass energies.  Since the soft anomalous dimension
matrix only depends on the ratio of the invariant cluster masses and
the cut off parameter $\mu$, it is possible to find an energy
independent prescription which should lead to the same distribution of
cluster masses.

A continuation of the arguments of the logarithms towards very small
cluster masses, $\ln(\frac{M_{\alpha \beta}^2}{\mu^2})\to
\ln(\frac{M_{\alpha \beta}^2}{\mu^2}+1)$ has been considered but did
not show any change in our findings and can as such be used to prevent
numerical instabilities should the relevant small masses be
encountered in a full model. The strong coupling parameter $\alpha_s$
is clearly a direct measure of the overall reconnection strength and
while we have chosen it to reflect the strong coupling at a lowish
scale, $\alpha_s=0.118$, it should, in practice also be considered a
tunable parameter of the model.

Until now we neglected quark masses completely, which lead to the
simplified form of the invariant cluster mass.  Assuming the same
quark (constituent) masses for light quarks ($0.3$ GeV for up and down
quarks) , which are used in the cluster hadronization model, only
small effects were found in terms of the mass distribution and no
sizeable effects on the reconnection probabilities.  For heavy quarks
($m_b \sim 4-5 \mathrm{GeV}$) more severe effects are expected but we
leave this topic for a detailed study in the scope of a full-fledged
model implementation. Switching off the Coulomb term does not change
our findings for the high-mass systems, while we see some effects for
small-mass systems.

\section{Towards a Full Model}
\label{sec:FullModel}

Due to its complexity, the approach followed in this paper is limited
to a small number of clusters and should be seen as a theoretical
consideration to constrain the structure of an improved colour
reconnection model.  It clearly makes a full colour flow evolution
un-feasible to implement on the typically large systems of clusters
encountered in a typical high energy collisions.  We will mainly use
the insights we gained from looking at evolution of small systems to
extrapolate a simplified model which could be suitable for
implementation. A crucial aspect is to identify independently evolving
subsystems in the cases with the highest number of clusters we did
consider here. This will allow us to formulate `microscopic' input to
a model iterating over small numbers of clusters within a large
ensemble.

\begin{figure}[t]
\centering
\includegraphics[width=7cm]{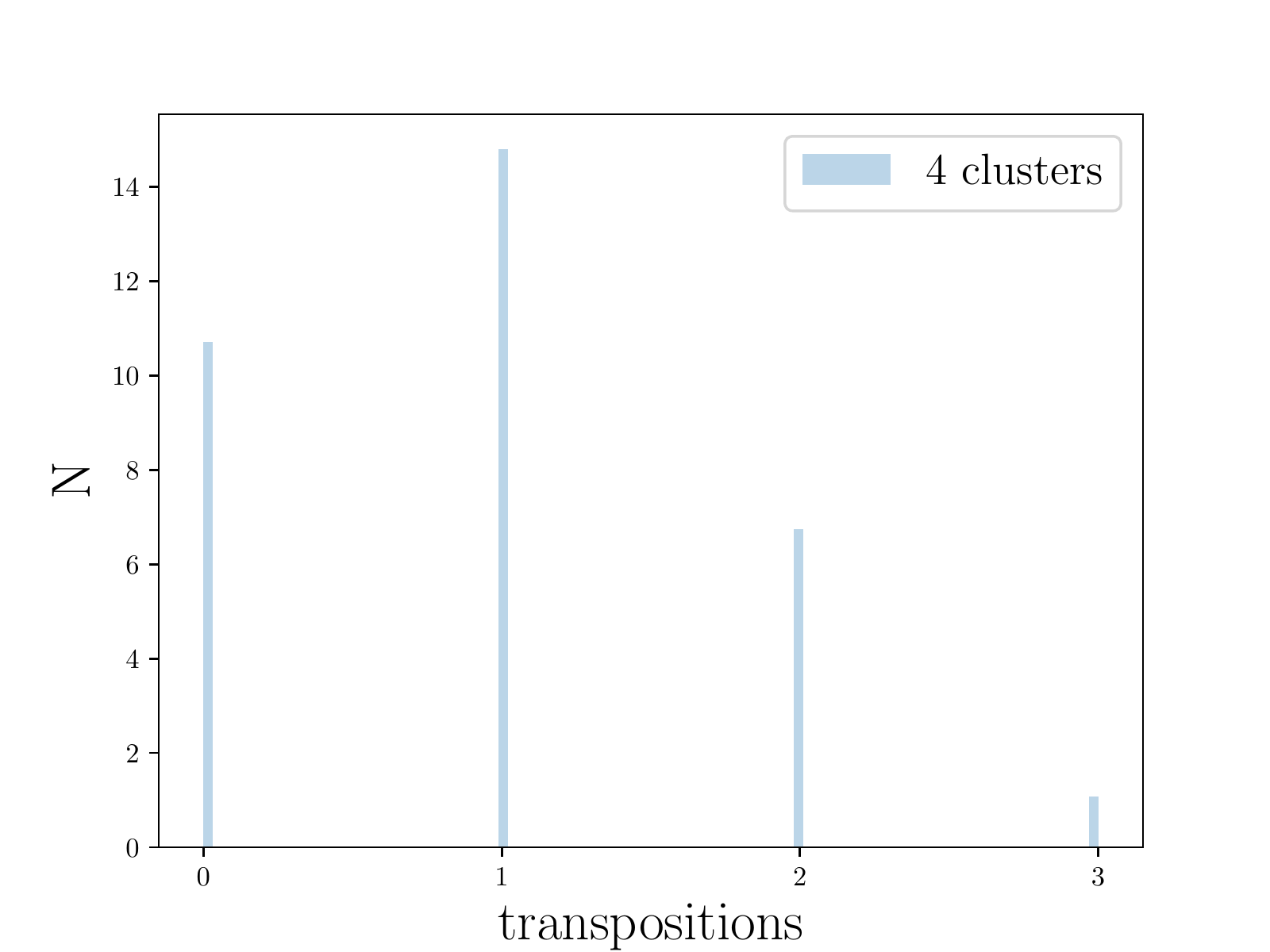}
\includegraphics[width=7cm]{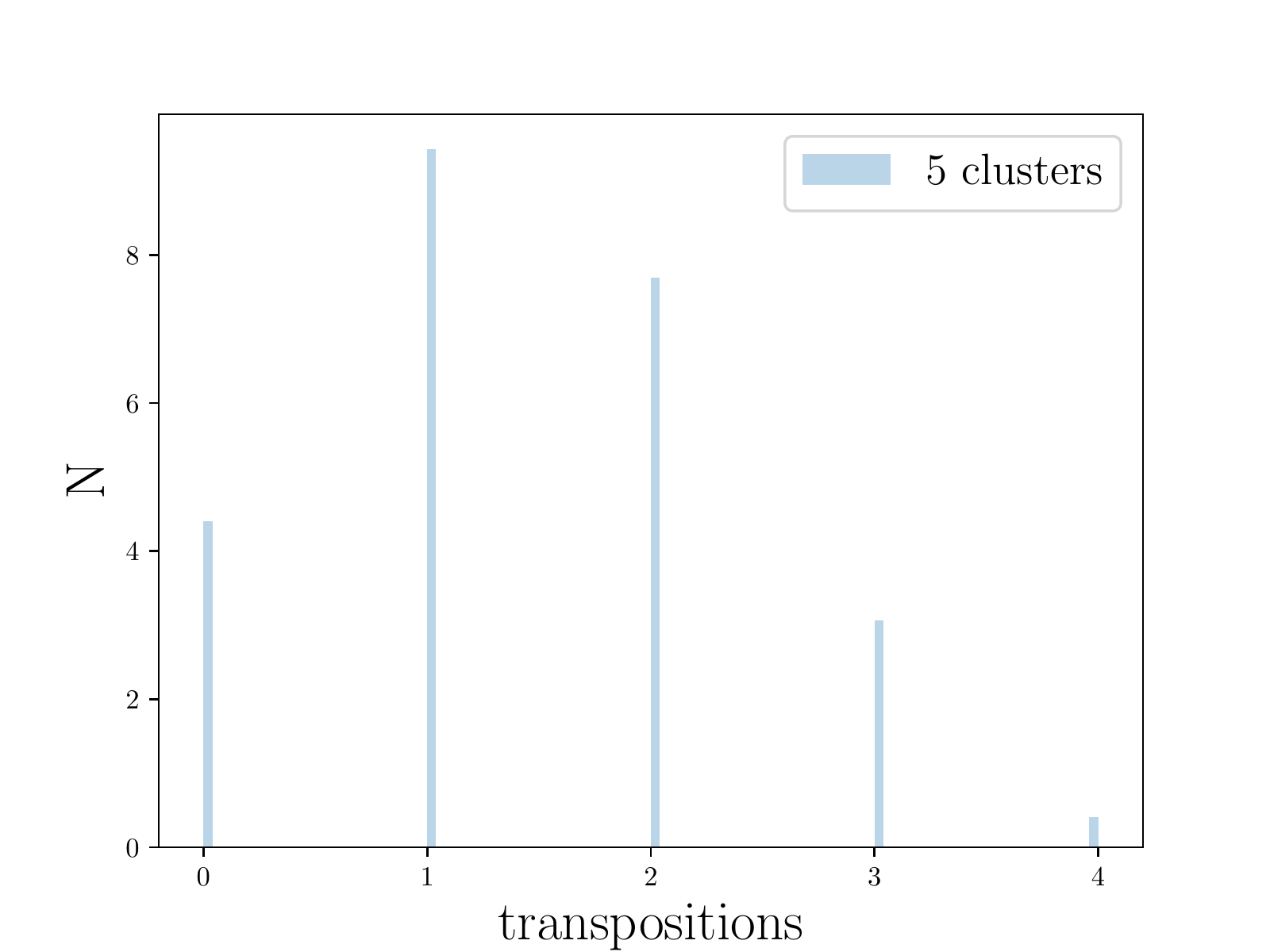}
\caption{ Number of transpositions between the initial state and the
  reconnected final state for four and five cluster evolution.}
\label{fig:Ntrans}
\end{figure}

\begin{figure}[t]
\centering
\includegraphics[width=7cm]{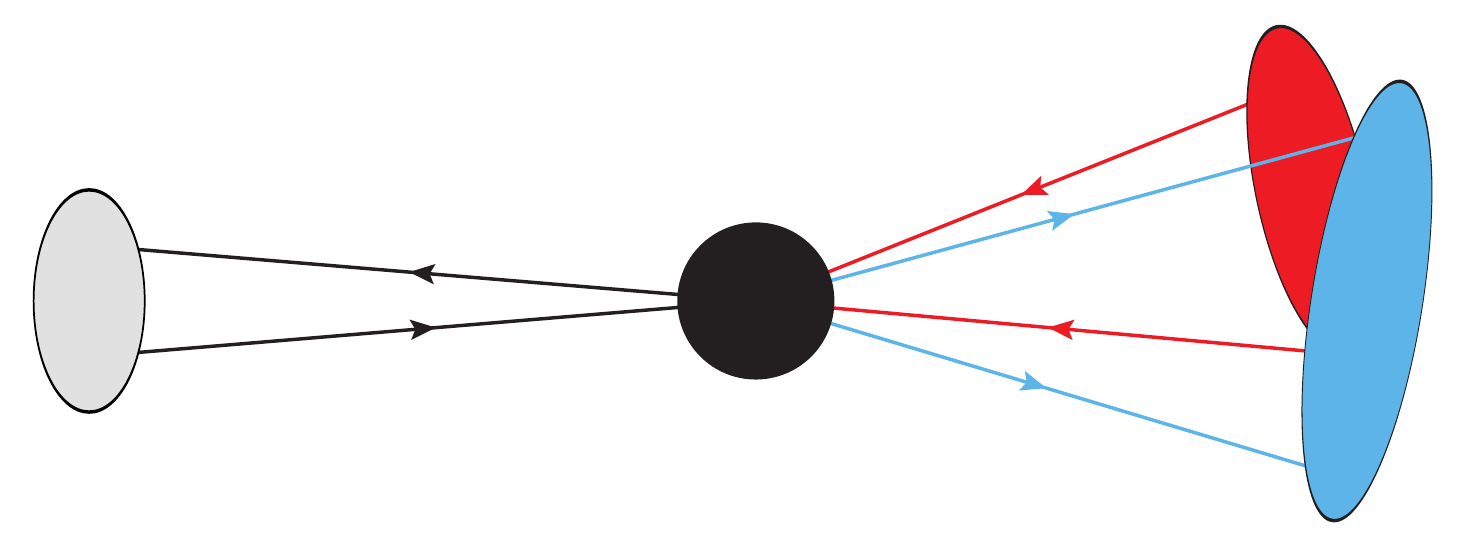}
\includegraphics[width=7cm]{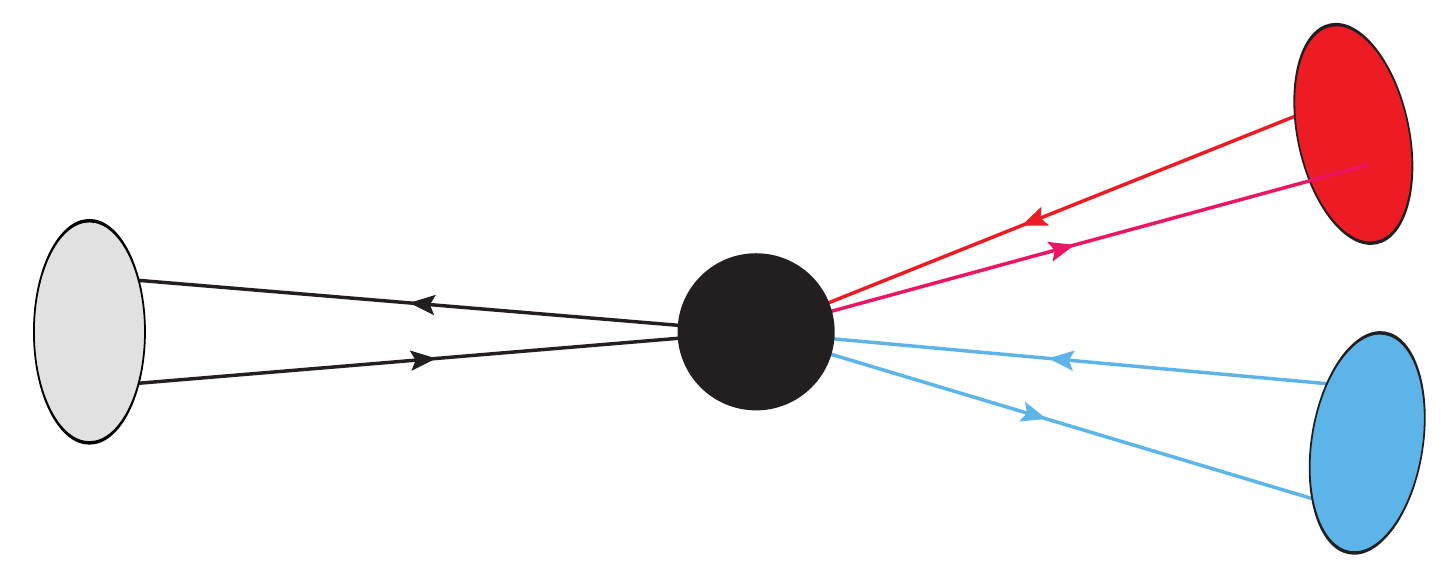}
\caption{ Sketch of a configuration where the two coloured clusters
  could be viewed as an independently evolving subsystem. Gluon
  exchanges which would lead to colour connections with the grey
  cluster are expected to be highly suppressed. Left: Initial colour
  flow. Right: Colour flow after reconnection.  }
\label{fig:subsystem2}
\end{figure}

The cluster configurations, being essentially colour structures in the
colour flow basis, can be labelled by permutations and an important
question to ask is what the minimum number of transpositions one
requires to transform the initial configuration into a final,
reconnected configuration. This number is directly related to the
power of the number of colours $N$ when evaluating the overlap between
the two colour structures, with more transpositions leading to a
higher $1/N$ suppression. In Fig.\,\ref{fig:Ntrans} we show the number
of transpositions for four and five cluster evolution where the phase
space was populated with the RAMBO method.
For both cases we note the peak at one transposition, {\it i.e.} a
reconnection within a two cluster system, between the initial and
final state.  This indicates the existence of independently evolving
subsystems where the contributions from the remainder of the event are
suppressed.  Crucially, this indicates that colour reconnection is not
simply a $1/N$ suppressed effect which would have indicated a much
higher rate of non-reconnected systems, as well as a much steeper drop
of the other reconnection dynamics with the number of transpositions.
Our finding is then also indicative of a choice of evolving small
subsystem out of a larger configuration as depicted in
Fig.\,\ref{fig:subsystem2}, where mixing with well separated clusters
can actually be neglected.

Colour reconnection models implemented in event generators often rely
on very simplified models in order to handle the complex structure of
hadronic collisions.  In an old model in the \texttt{Herwig} event
generator, for example, reconnections would only be accepted if they
allowed for a smaller mass configuration, and with a fixed probability
which essentially was inferred by tuning to underlying event data.
While this approach has benefits in terms of efficiency and simplicity,
and has shown to provide a reasonable description of data
\cite{Gieseke:2012ft} it does not take into account the full kinematic
dynamics and complexity of a hadronic event.  

\begin{figure}[t]
\centering
\includegraphics[width=7cm]{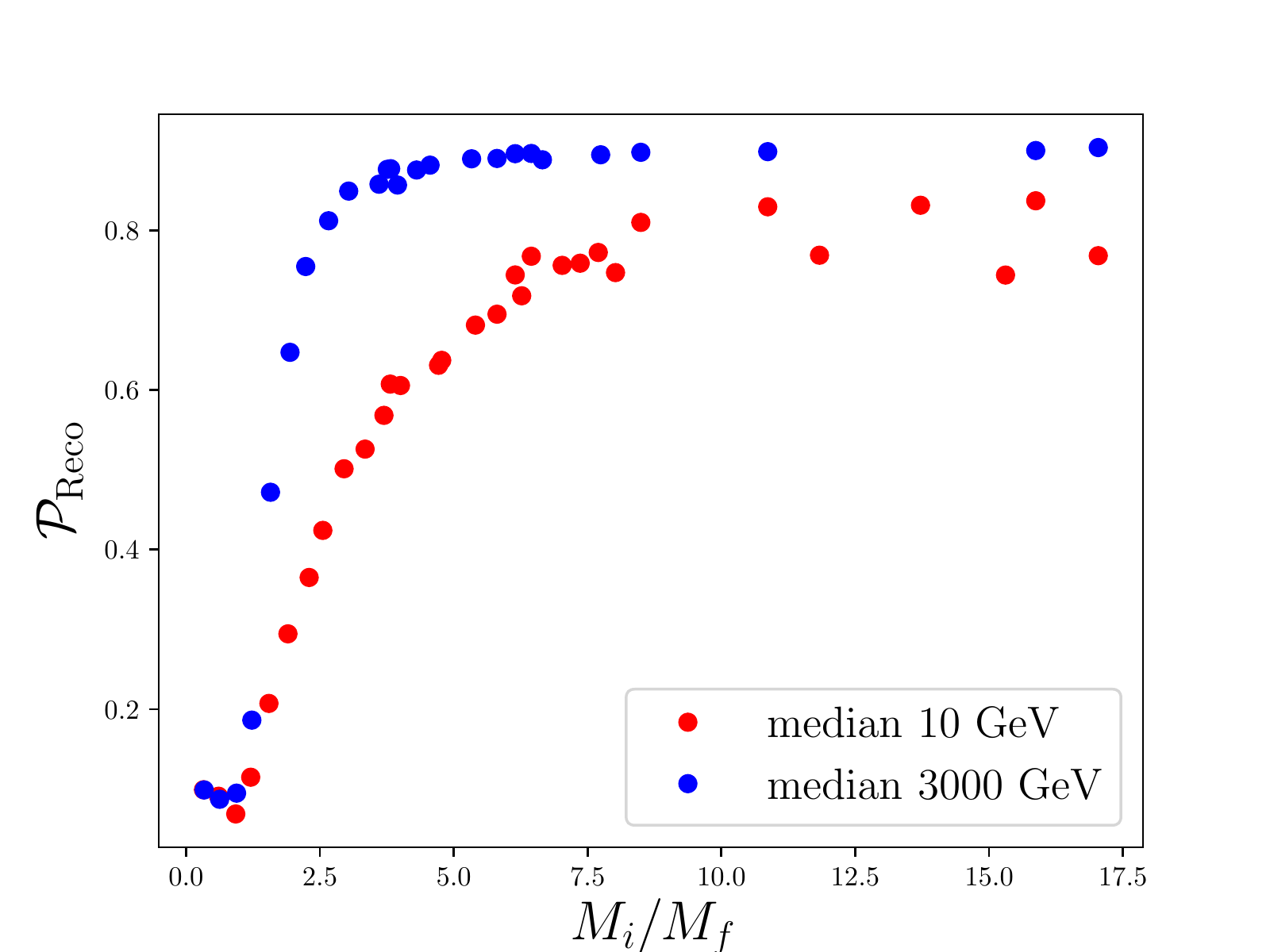}
\caption{ Parametrisation of the reconnection probability for two
  cluster evolution in terms of the ratio of invariant cluster masses
  $M_i/M_f$,  where $M_i$ is the sum of invariant cluster masses of the initial
colour flow and $M_f$ is the sum of invariant cluster masses of the
alternative colour flow of a two cluster system.   }
\label{fig:pRecoParam}
\end{figure}

In order to make contact with this simple model, and to highlight the
fact that geometric models as well as the non-trivial kinematic
dependence of our {\it Ansatz} provide a much more dynamic model, we
consider the reconnection probability projected to the variable of the
old model, which essentially is the ratio of the sum of cluster masses
before and after reconnection. Having generated kinematics of two
clusters with the RAMBO algorithm with $\sqrt{s}=10\,\mathrm{GeV}$,
$\mu=0.01\,\mathrm{GeV}$, $\alpha_s=0.118$ and for comparison with
$\sqrt{s}=3000\,\mathrm{GeV}$, $\mu=1\,\mathrm{GeV}$, $\alpha_s=0.118$
to visualise the energy dependence, and plot the median of the
reconnection probabilities over the ratio of the sum of invariant
cluster masses, which would result from the two different possible
colour flows, see Fig.\,\ref{fig:pRecoParam}.
The old model would
here only have put a step function in place, with no further kinematic
dependence present. We also stress the fact that the reconnection
probability does not vanish, but saturates at $\approx 0.9$ if
reconnection would result in a lower sum of invariant cluster masses
and mounts up at $\approx 0.1$ if $M_i=M_f$ or $M_i<M_f$.  These
limits can already be obtained from our analytic studies in
Sec.\,\ref{sec:TwoClusterSandbox}. For both energies it is clear that
the dependence of the reconnection probability on the ratio of cluster
masses is a more dynamical one than a simple step function.

We therefore suggest to indeed use our findings as an input to more
sophisticated reconnection models along the lines of \cite{Gieseke:2017clv}:
\begin{itemize}
\item The analytically known reconnection probability from the
  evolution of two cluster systems can, with the finding of mostly
  independently evolving two cluster systems, be used directly to
  improve the model assumption of these type of reconnections.
\item The fact that also three cluster systems seem to be rather
  detached in a large ensemble can be used to supplement the baryon
  production mechanism in \cite{Gieseke:2017clv} with a more dynamic
  reconnection probability, possibly based on approximating this
  evolution to the first few orders in a $1/N$ expansion
  \cite{Platzer:2013fha}.
\item The baryonic reconnection mechanism, which has so far not
  considered the possibility of un-connecting baryonic clusters in an
  evolution picture or statistical model \cite{Gieseke:2012ft}.
\end{itemize}
All of these mechanisms are contained within our approach and that
such a model will be highly predictive in the sense that with the
analogue of the strong coupling $\alpha_s$ and the soft scale $\mu$ it
contains effectively two, possibly three parameters if one wants to
include the number of colours $N$, as well. We finally note that,
though we have essentially been considering the cluster hadronization
model in our considerations, a similar dynamics could be implemented
in a string picture.

\section{Conclusions and Outlook}
\label{sec:Conclusions}

We have studied to what extent the structure of perturbative colour
evolution can be used as an input to improve or constrain existing
colour reconnection models. In particular, we have analytically solved
the evolution of a two-cluster system, as well as numerically studied
the evolution of larger systems of up to five clusters. We have found
that there is indeed a highly dynamic and non-trivial re-arrangement
of colour structures already from a simple {\it Ansatz} using a
one-loop soft anomalous dimension, which confirms earlier work on
geometrically inspired reconnection models \cite{Gieseke:2017clv}.

The full evolution in colour space is, however, not feasible in a
realistic model which needs to cope with several tens to hundreds of
clusters. However we have found evidence that in the evolution of
larger systems the bulk of the reconnection effects is isolated in
small subsystems of two to three clusters which allow to build an
iterative model, which can also be based on approximations of the
evolution operator \cite{Platzer:2013fha}. Our framework also allows
to include a probability of re-connecting baryonic clusters into
mesonic systems, an important aspect which has not been considered in
models of baryonic reconnection so far, but could result in a detailed
balance mechanism to establish a realistic fraction of baryons for
given partonic constituent dynamics.

We stress the fact that our approach should not only be considered as
a motivation for improved models of non-perturbative colour
reconnection but does highlight that the perturbative mixing of colour
structures, mediated through virtual soft gluon exchanges, should be
considered an important ingredient in new approaches to improving
parton shower algorithms beyond the leading-$N$ level
\cite{Martinez:2018ffw}, and that these effects are in general not
mediated in a probabilistic manner, or through tree-level amplitudes.
However until such algorithms are fully available, and the dynamics of
hadronization is understood in this context, we postpone further
aspects to future work and use the findings obtained here as an
improved input to existing colour reconnection models.

\section*{Acknowledgments}
This work has been supported in part by the BMBF under grant number 05H15VKCCA
and 05H18VKCC1.  This work was also supported by the MCnetITN3 H2020 Marie
Curie Initial Training Network, contract number 722104, as well as the
European Union's Horizon 2020 research and innovation programme (grant
agreement No 668679), and the COST action (``Unraveling new physics at the LHC
through the precision frontier'') No.\ CA16201. S.P. is grateful to KIT, CERN
and MITP for their kind hospitality, and P.K. is grateful to Universit\"at
Wien for their hospitality, while several aspects of the present work have
been addressed. P.K. also acknowledges the support received from the Karlsruhe
House of Young Scientists. A.S. acknowledges support from the National Science
Centre, Poland Grant No. 2016/23/D/ST2/02605 and the grant 18-07846Y of the
Czech Science Foundation (GACR).

\bibliography{colour-reconnection}

\end{document}